\documentclass[11pt]{article}
\usepackage{amssymb}
\usepackage{citesort}
\usepackage{graphics,graphicx}

\renewcommand{\theequation}{\arabic{section}.\arabic{equation}}

\begin{document}
\begin{titlepage}

\begin{flushright}
hep-th/0603164\\
BONN--TH--2006--2 \\
 IFT UWr 0103/006
\end{flushright}

\vskip 2cm

\begin{center}
{\LARGE\bf\sf Semiclassical limit of the FZZT Liouville theory}
\end{center}

\bigskip

\begin{center}
    {\large\bf\sf
    Leszek Hadasz}\footnote{Alexander von Humboldt Fellow;
    e-mail: hadasz@th.if.uj.edu.pl} \\
\vskip 1mm
    Physikalisches Institut\\
    Rheinische Friedrich-Wilhelms-Universit\"at \\
    Nu\ss{}allee 12, 53115 Bonn, Germany
    \\
    and \\
    M. Smoluchowski Institute of Physics, \\
    Jagiellonian University, \\
    W.\ Reymonta 4, 30-059 Krak\'ow, Poland \\
\vskip 5mm
    {\large\bf\sf
    Zbigniew Jask\'{o}lski}\footnote{e-mail: jask@ift.uni.wroc.pl
}\\

\vskip 1mm
   Institute of Theoretical Physics\\
 University of Wroc{\l}aw\\
 pl.\ M.\ Borna 9, 50-204 Wroc{\l}aw,
 Poland
\end{center}

\vskip 1cm

\begin{abstract}
The semiclassical limit of the FZZT Liouville  theory
on the upper half plane with bulk operators of arbitrary type
and with elliptic boundary operators is analyzed.
We prove the Polyakov conjecture for an
appropriate classical Liouville action.
This action is calculated in a number of cases:
one bulk operator of arbitrary type,
one bulk and one boundary,
and  two boundary elliptic operators.
The results are in agreement with the classical
limits of the corresponding quantum correlators.
\end{abstract}

\end{titlepage}

\section{Introduction}

In the last few years the quantum Liouville theory
attracted again  a considerable attention, mainly for its application
in describing
D-brane dynamics in non-compact curved or time dependent backgrounds
(for review and  references see \cite{Nakayama:2004vk}).
This renewed interest was supported by a
recent progress in
the quantum Liouville theory on bordered surfaces,
where exact analytic forms of all basic correlators were derived
\cite{Teschner:2000md,Fateev:2000ik,Zamolodchikov:2001ah,Ponsot:2001ng,Hosomichi:2001xc,Ponsot:2003ss}
 by powerful conformal bootstrap
methods. These methods were previously successfully applied in rederiving \cite{Teschner:1995yf}
 the DOZZ bulk three point function \cite{Dorn:1994xn,Zamolodchikov:1995aa}
and proving its consistency
\cite{Ponsot:1999uf,Ponsot:2000mt,Teschner:2001rv}.

In the case of the DOZZ three point function on the sphere the exact
result agrees with the perturbative calculations \cite{Thorn:2002am}
based on the hamiltonian approach
\cite{Braaten:1982yn,Braaten:1983np,Braaten:1983qs}. A perturbative
check of the exact formula for the one point function on the
pseudosphere was given in
\cite{Zamolodchikov:2001ah,Menotti:2003km}. This suggests that the
bootstrap solution should have a well defined path integral
representation. In all the cases in which a classical solution
exists one then  could expect that the quantum expressions arise as
a result of integrating fluctuations of classical background
geometries.

In the case of ``heavy'' elliptic weights on compact surfaces
a systematic  functional formulation of Liouville theory
was first proposed long time ago
by Takhtajan \cite{Takhtajan:yk,Takhtajan:1994vt} (see \cite{Takhtajan:2005md}
for recent generalizations). This so called geometric approach
has been recently compared with the exact bootstrap solution on the pseudosphere
\cite{Menotti:2004uq}. It was shown in particular that the choice of regularization
is crucial for the agreement with the bootstrap approach. Recently a
functional representation of Liouville correlators with heavy elliptic charges
on the sphere \cite{Menotti:2004xz}, the pseudosphere \cite{Menotti:2006gc},
and the disc \cite{Menotti:2006tc} has been developed. This representation  agrees
with the bootstrap solution at least up to one loop calculations.

In the present paper we address the problem of calculating the classical limit of
the FZZT Liouville theory for heavy charges. Our motivation is twofold.
In spite of the recent progress in constructing the  path integral
representation of Liouville correlators for elliptic weights there is still
an open and interesting problem
of factorization in functional approach. Due to the global
character of the Liouville action related to its specific dependence on
the background metric
 it is not simply described by the
standard cutting-open procedure.
It seems that in order to handle this problem one should
first construct
 a functional  representations for
 Liouville correlators with hyperbolic weights.
The FZZT theory provides simple cases where
 exact solutions are known and both problems can be relatively easily analyzed.
The classical limit is just the first step of such calculations.

The second, probably more important motivation is the semiclassical limit itself.
It turned out that analyzing the quantum Liouville theory in this limit one gets
essentially new insight into the classical hyperbolic geometry.
One of the first results of this type
was the so called Polyakov conjecture,
originally obtained  as a classical limit of the Ward identity
and proved latter as an exact theorem
\cite{ZoTa1,ZoTa2,Zograf,Cantini:2001wr,Takhtajan:2001uj,Hadasz:2003kp}.
It says that  the classical Liouville action is a generating
function for the  accessory parameters of the Fuchsian
uniformization of the punctured sphere.

More intriguing results are related with the classical conformal block
 $\,f_{\delta}
\!\left[_{\delta_{4}\;\delta_{1}}^{\delta_{3}\;\delta_{2}}\right](x),$ defined
by the classical limit  of the BPZ quantum conformal block \cite{Belavin:1984vu}
with heavy weights $\Delta=Q^2\delta,$
$\Delta_i=Q^2\delta_i,$
\cite{Zamolodchikov:1995aa,Zam0,Zam}:
\begin{equation}
\label{defccb}
{\cal F}_{\!1+6Q^2,\Delta}
\!\left[_{\Delta_{4}\;\Delta_{1}}^{\Delta_{3}\;\Delta_{2}}\right]
\!(x)
\; \sim \;
\exp \left\{
Q^2\,f_{\delta}
\!\left[_{\delta_{4}\;\delta_{1}}^{\delta_{3}\;\delta_{2}}\right]
\!(x)
\right\}.
\end{equation}
It was argued in \cite{Zamolodchikov:1995aa}
that the 4-point Liouville classical action can be expressed in terms of
3-point actions and the classical conformal block in
a given channel calculated
for a saddle point value of the intermediate weight.
Since there are three  different decompositions
of the 4-point action one gets consistency conditions
called the classical bootstrap equations \cite{Hadasz:2005gk}.
It was also conjectured \cite{Hadasz:2005gk} that the saddle point
weight is closely related to the length of the
closed geodesic in the corresponding channel.
These statements are far from being proved in a rigorous way,
but there are many nontrivial numerical checks instead \cite{Hadasz:2005gk}.
Let us finally stress that once the 4-point Liouville
action is known one can in principle calculate the
uniformization of the 4-punctured sphere \cite{HadJas}, which is
a long standing open mathematical problem.
Analyzing various classical limits one may hope to
gain some new information on the classical
conformal block, which up to now is only available
 through  term by term symbolic calculations from  the limit (\ref{defccb}).

The organization of the paper is as follows.
In Sect.\ 2, following the ideas of \cite{Bilal:1987cq},
we formulate the $SL(2,\mathbb{R})$-monodromy problem with boundary.
In Sect.\ 3 we introduce
 an appropriate classical Liouville action and derive
 the formulae for its partial derivatives
 with respect to bulk and boundary conformal weights,
 and bulk and boundary cosmological constants. A novel technical result
is the formula for the partial derivative of the action
 with respect to the ratio $\omega_j ={m_j\over \sqrt{m}}$
 in terms of special values of some conformal map.
 Details of its derivation are explained in the Appendix.

In Sect.\ 4 we present  a proof of  the Polyakov conjecture for elliptic
boundary weights and arbitrary bulk weights. This extends the previous results
\cite{ZoTa1,ZoTa2,Zograf,Cantini:2001wr,Takhtajan:2001uj,Hadasz:2003kp}
to the FZZT Liouville theory and is one of the main results of this paper.

The last section contains four examples of explicit calculations
of the classical actions both from classical solutions and from the
classical limit of  exact quantum expressions.
The simplest one is the case of one bulk elliptic singularity
considered in Subsect.\ 5.1. The case of one bulk hyperbolic
singularity is considered in Subsect.\ 5.2. This calculations
provide an additional support for the construction
of the classical action for hyperbolic singularities
proposed and analyzed in \cite{Hadasz:2003kp,Hadasz:2003he}.
The cases of two boundary, and one bulk one boundary
elliptic weights are calculated in Subsect.\ 5.3 and Subsect.\ 5.4,
respectively.
Both ways of calculation the classical Liouville action
are rather complicated in these cases.
The full agreement of the results
provides an additional  strong evidence
that the classical limit of the DOZZ bootstrap solution
exists and is properly described by a classical action
satisfying the equations derived in Sect.\ 3 and 4.

\section{The monodromy problem with boundary}
\setcounter{equation}{0}

Let us consider the Fuchs equation
\begin{equation}
\label{Fuchs}
\partial^2\psi(z) + T(z)\psi(z) = 0,
\end{equation}
with the energy--momentum tensor of the form
\begin{equation}
\label{T}
T(z) =
 \sum\limits_{i=1}^{m}\left[
 \frac{\delta_i}{(z-z_i)^2} +\frac{\delta_i}{(z-\bar z_i)^2} +
\frac{c_i}{z-z_i}+\frac{\bar c_i}{z-\bar z_i}\right]
+
\sum\limits_{j=1}^{n}\left[\frac{\delta^{\rm \scriptscriptstyle B}_j}{(z-x_j)^2} +
\frac{c^{\rm \scriptscriptstyle B}_j}{z-x_j}\right],
\end{equation}
where $\delta_i$ are bulk conformal weights located at the points $z_1,\dots,z_m$ of the upper half-plane,
$\delta^{\rm \scriptscriptstyle B}_j$ are boundary conformal weights located at the points $x_1< \dots < x_n$ of the real axis,
and $c_i, c^{\rm \scriptscriptstyle B}_j$ are accessory parameters. We assume that  $\delta_i, \delta^{\rm \scriptscriptstyle B}_j, c^{\rm \scriptscriptstyle B}_j$ are real
so that
\begin{equation}
\label{tbar}
\overline{T(z)} = T(\bar z).
\end{equation}
The requirement that $T(z)$ is regular at the infinity
implies the relations
\begin{eqnarray*}
\label{restrictions}
2\sum\limits_{i=1}^{m}\Re c_i+  \sum\limits_{j=1}^{n}c^{\rm \scriptscriptstyle B}_j &=& 0, \\
2\sum\limits_{i=1}^{m} \Re (z_ic_i)+  \sum\limits_{j=1}^{n}x_jc^{\rm \scriptscriptstyle B}_j &=& -
2\sum\limits_{i=1}^{m}\delta_i - \sum\limits_{j=1}^{n}\delta^{\rm \scriptscriptstyle B}_j,  \\
2\sum\limits_{i=1}^{m} \Re (z_i^2c_i)+  \sum\limits_{j=1}^{n}x_j^2c^{\rm \scriptscriptstyle B}_j &=&-
4\sum\limits_{i=1}^{m}\Re z_i \delta_i - 2\sum\limits_{j=1}^{n}x_j \delta^{\rm \scriptscriptstyle B}_j.
\end{eqnarray*}
A fundamental systems of solutions
\(
\Psi=\left[
\begin{array}{r}
\psi^-\\
\psi^+
\end{array}
\right]
\)
to the Fuchs equation (\ref{Fuchs}) is normalized if
$$
\psi^-\partial \psi^+ -\psi^+ \partial \psi^-  = 1.
$$
Let $L_j$ denote  the part of the real axis between $x_{j}$
and $x_{j+1}$ (with the exception of $L_n$ denoting the set of points on the real axis to
the right of $x_n$ and to the left of $x_1$). It follows from (\ref{tbar}) that
for each boundary segment $L_j$
there exist normalized solutions
$
\Psi_j
$
 to the Fuchs equation
(\ref{Fuchs})  regular and real along $L_j$. For any other normalized solution $
\Psi
$
we define the matrices
$$
M_j =\Sigma^{-1}B^T_j \,\overline{ \Sigma}\, \overline{B}_j, \;\;\;j=1,\dots,n,
$$
where
$
\Sigma=\left[
\begin{array}{rr}
0&1\\
-1&0
\end{array}
\right]
$
and
the matrix $B_j$ is determined by the relation
$
\Psi
=
B_j
\Psi_j.
$

We are interested in  the following
version of the Riemann-Hilbert problem \cite{Bilal:1987cq}.
For given sets of positive weights $\{\delta_i\}_{i=1}^m$, $\{\delta^{\rm \scriptscriptstyle B}_j\}_{j=1}^n$
and real numbers $\{\omega_j\}_{j=1}^n$
one has to adjust
the accessory parameters in such a way that the Fuchs equation
(\ref{Fuchs})
admits a normalized fundamental system
$
\Psi
$
 of solutions, such that:
\begin{enumerate}
\item
monodromies around all  singularities $z_i$ of the upper half plane belong to $SL(2,\mathbb{R})$;
\item
the function $-i \Psi^T \cdot\Sigma \cdot\overline{\Psi} $ is strictly
positive or strictly negative on the upper half plane  except
the points $z_i,\ i=1,\dots,m$ and $x_j,\ j=1,\dots,n;$
\item
for each boundary segment $L_j$ the  boundary condition
$$
{\rm Tr}\,M_j= \left\{
\begin{array}{llll}
+\omega_j&{\rm if}&-i \Psi^T \cdot\Sigma \cdot\overline{\Psi}\, >\, 0,\\
-\omega_j&{\rm if}&-i \Psi^T \cdot\Sigma \cdot\overline{\Psi}\, <\, 0,
\end{array}
\right.
$$
 is satisfied.
\end{enumerate}
If $\Psi$ is a solution to the monodromy problem above
then  the relation
\begin{equation}
\label{general:solution}
{\rm e}^{-{\varphi}} =
 \left(\frac{\sqrt m}{2i} \Psi^{\rm T}(z)\cdot\Sigma
\cdot \overline{\Psi( z)}\right)^2
\end{equation}
determines  a conformal factor $\varphi$ on the upper half plane
satisfying the Liouville equation
\begin{equation}
\label{bulk}
\partial\bar\partial \varphi = \frac{m}{2}{\rm e}^{\varphi}
\end{equation}
and the boundary conditions
\begin{equation}
\label{boundary2}
\left.\partial_y {\rm e}^{-{\varphi\over 2 }}\right|_{Lj} =
-{\textstyle {\sqrt{m}\over 2}}\omega_j,
\hskip 5mm
j=1,\dots,n.
\end{equation}
They are usually written in the form
\begin{equation}
\label{boundary1}
\left.\partial_y\varphi\right|_{L_j} = m_j{\rm e}^{{\varphi\over 2 }},
\hskip 5mm
 m_j= \omega_j\sqrt{m}\, ,
\end{equation}
where $m_j$ are so called boundary cosmological constants.
 Note that $T(z)$ of (\ref{T}) is the
classical energy momentum tensor of the solution $\varphi$:
\begin{equation}
\label{Tclass}
T(z) =T_{\rm cl}(z)=  -{1\over 4} (\partial \varphi)^2
+{1\over 2} \partial^2\varphi=
-{\rm e}^{{\varphi\over 2 }}\partial^2 {\rm e}^{-{\varphi\over 2 }}.
\end{equation}

The conformal factor $\varphi$ is a regular single valued function on the upper half plane $\Im z\geqslant 0$
except the singular points $z_i$, $x_j$.
It defines the hyperbolic metric ${\rm e}^\varphi dzd\bar z$
with the constant negative scalar curvature
$$
R= -{\rm e}^{-\varphi} \Delta \varphi = -4{\rm e}^{-\varphi} \partial \bar \partial \varphi = - 2m,
$$
and with the constant geodesic curvature of each boundary sector $L_j$:
$$
\kappa_j=
\left.
{\textstyle{1\over 2}}
{\rm e}^{-{\varphi\over 2 }}
n^a\partial_a\varphi
\right|_{Lj}
= -\frac{m_j}{2}
 =- \frac{\sqrt{m}}{2}\omega_j.
$$

\section{The classical Liouville action}
\setcounter{equation}{0}

In order to simplify our considerations we assume that all weights are elliptic:
$$
\delta_i= {1-\xi^2_i\over 4},\;\;j=1,\dots,m,
\hskip 5mm
\delta^{\rm \scriptscriptstyle B}_j= {1-\nu^2_j\over 4},\;\;j=1,\dots,n,
\hskip 5mm
0 < \xi_i,\nu_j < 1.
$$
It follows from (\ref{T}), (\ref{tbar}), and (\ref{Tclass}) that
 the most singular terms in the expansions of the
Liouville field around the locations of elliptic weights  read
\begin{equation}
\label{singularity}
\varphi(z,\bar z) \sim -2(1-\xi_i)\log|z-z_i|,
\hskip 5mm
\varphi(z,\bar z) \sim -2(1-\nu_j)\log|z-x_j|.
\end{equation}
Let $X$ be the upper half plane with the discs of radii $\epsilon$ around
the points $z_i,\; i = 1,\ldots,m$ and the semi-discs of radii
$\epsilon$ around the points
$x_j,\; j = 1,\ldots,n$ removed. Denote by $S_i$ the boundary of the disc
around $z_i$ and by $s_j$ the semicircle forming a boundary of the semi-discs around
$x_i.$  Finally, let $l_i$ denote the part of the real axis between $x_{i}+\epsilon$
and $x_{i+1}-\epsilon$ (with the exception of $l_n$ denoting the set  $[-R,x_1-\epsilon]\cup [x_n+\epsilon,R]$).
The regularized action functional is defined by
\begin{eqnarray}
\nonumber
S[\phi] & = & \lim_{\epsilon \to 0} S_{\epsilon}[\phi], \\
\nonumber
S_\epsilon[\phi] & = & \lim_{R \to \infty } S_{\epsilon}^R[\phi], \\
\label{action}
S_{\epsilon}^R[\phi] & = &
\frac{1}{4\pi}\int\limits_{X}\!d^2z\left[\partial\phi\bar\partial\phi + m{\rm
    e}^{\phi}\right]
\\
\nonumber
&+&
\sum\limits_{i=1}^m\left[
\frac{1-\xi_i}{4\pi}\int\limits_{S_i}\kappa_z|dz|\ \phi
-{(1-\xi_i)^2\over 2} \log\epsilon\right]
\\
\nonumber
&+& \sum\limits_{j=1}^n
\frac{m_j}{4\pi}\int\limits_{l_j}dx\ {\rm e}^{\phi\over 2}
+\sum\limits_{j=1}^n\left[
\frac{1-\nu_j}{4\pi}\int\limits_{s_j}\kappa_z|dz|\ \phi
-{(1-\nu_j)^2\over 4}\log\epsilon\right]
\\
\nonumber
&+&
\frac{1}{2\pi R}\int\limits_{s_R}\!|dz|\;\phi
+ \log R,
\end{eqnarray}
where $s_R$ is the semicircle $|z|=R$ on the upper half plane.
The action is constructed in such a way that
 for fields satisfying
$$
\begin{array}{lll}
\phi \sim -2(1-\xi_i)\log|z-z_i|\; &\; {\rm for} \;&\; z \to z_i,\\
\phi \sim -2(1-\nu_j)\log|z-x_j|\;&\; {\rm for} \;&\; z \to x_j,\\
\phi \sim -4\log|z|\;&\; {\rm for} \;&\; z \to \infty,
\end{array}
$$
 the limit in
(\ref{action}) exists and the equation $\delta S[\phi] = 0$ gives
(\ref{bulk}) and (\ref{boundary1}).

Let $\varphi(z,\bar z)$ denote a solution of (\ref{bulk}) and (\ref{boundary1}) with some
specified values of $m,\xi_i,z_i$ and $m_j, \nu_j, x_j$. We define the classical Liouville action
$S_{\rm cl}$ as the value of the action functional
(\ref{action}) calculated on this solution. Using  the
equations of motion and the boundary condition satisfied by $\varphi(z,\bar
z)$ one immediately gets
\begin{eqnarray}
\label{der_xi}
\frac{\partial S_{\rm cl}}{\partial\xi_i}
& = & \lim_{\epsilon\to 0}\left[-\frac{1}{4\pi}\int\limits_{S_i}\kappa_z|dz|\ \varphi
  +(1-\xi_i)\log\epsilon\right], \\
  \label{der_nu}
\frac{\partial S_{\rm cl}}{\partial\nu_j}
& = & \lim_{\epsilon\to 0}\left[-\frac{1}{4\pi}\int\limits_{s_j}\kappa_z|dz|\ \varphi
  +{1-\nu_j\over 2}\log\epsilon\right].
\end{eqnarray}
Shifting the classical solution $\varphi = \widetilde \varphi - \log m$
one obtains
\begin{eqnarray}
\label{der_m}
S_{\rm cl}(m,\xi_i,z_i,m_j, \nu_j, x_j)&=&
S_{\rm cl}(1,\xi_i,z_i,{\textstyle {m_j\over \sqrt{m}}}, \nu_j, x_j)
\\ \nonumber
&+&
\left(
\sum\limits_{i=1}^m{1-\xi_i\over 2}
+
\sum\limits_{j=1}^n{1-\nu_j\over 4}-{1\over 2}
\right)\log m  .
\end{eqnarray}

It is convenient to regard the classical action as a function of the
variables $m$ and $\omega_j$ (instead of $m$ and $m_j$). One than has
\begin{equation}
\label{der_omega}
\frac{\partial S_{\rm cl}}{\partial\omega_j}
=
{1\over 4\pi} \int\limits_{L_j} dx\, {\rm e}^{\widetilde \varphi \over 2},
\hskip 5mm
j=1,\dots,n.
\end{equation}

Equations (\ref{der_xi}) and (\ref{der_nu}) express
the derivatives of the classical
action in terms of the  classical metric in the
vicinity of the singularities.
It is useful to work out  a similar ``local'' expression
for the integral in the r.h.s.\ of (\ref{der_omega}).
To this end note that the solution $\Psi$ to the monodromy problem
described in the previous section
determines a multivalued  analytic function from
$$
{\cal M}=\{z\in \mathbb{C}:\Im z\geqslant 0\}\cup\{\infty\}\setminus
\{z_1,\dots,z_n,x_1,\dots,x_j\}
$$
to the upper half plane $\mathbb{H}=\{z\in \mathbb{C}:\Im z> 0\},$
\begin{equation}
\label{map:rho}
{\cal M} \, \ni\, z \;\; \mapsto \;\;\rho(z)
= \frac{\psi^{-}(z)} {\psi^{+}(z)}\,\in\, {\mathbb H}.
\end{equation}
The  pull-back of the standard
hyperbolic metric on $\mathbb H$ by any branch of $\rho$ yields a
regular hyperbolic metric on $\cal M$:
\begin{equation}
\label{pullback}
{\rm e}^{\varphi(z,\bar z)}dz d\bar z = \frac{1}{m \left(\Im\,\rho\right)^2} d\rho d\bar\rho .
\end{equation}
One can always chose a branch of $\rho$ such that the image $\rho(L_j)$ of the
boundary segment $L_j$ is a connected open  curve in $\mathbb{H}$
joining the points
$$
\rho_j = \lim_{x\to x_j +} \rho(x),
\hskip 5mm
\rho_{j+1} =\lim_{x\to x_{j+1} -} \rho(x).
$$
This curve has the constant geodesic curvature $-{\omega_j\over 2}$ with respect
to the Poincar\'{e} metric on $\mathbb{H}$ (the sign being determined
with respect to $\rho(\cal M)$).
It
 admits a simple description on the Lobachevsky plane as the arc of the Euclidean
circle containing the points $\rho_j, \rho_{j+1}$ and with its radius $R$ and its center $\cal O$
determined by the condition that the Euclidean distance of $\cal O$
from the real axis is equal to
$
|\frac{\omega_j}{2}|\, R
$.
For $|\omega_j|\leqslant 2$ there are two and  for $|\omega_j|> 2$
four  different  arcs with this property.
It is shown in the Appendix that in both cases the hyperbolic length
depends only on the location of the endpoints
$\rho_j, \rho_{j+1}$ and $|\omega_j|$.
It follows from (\ref{pullback}) that the length of the boundary component
$L_j$ with respect to the metric ${\rm e}^{\varphi(z,\bar z)}$ is equal to
the length of its image $\rho(L_j)$ in the Lobachevsky plane. Using the
explicit expressions (\ref{app:1}) and (\ref{app:2}) one gets
\begin{eqnarray}
\label{deromegafinal:1}
\frac{\partial S_{\rm cl}}{\partial\omega_j}
& = &
\frac{1}{\pi\sqrt{4-\omega_j^2}}\
{\rm arcsinh}\left[\sqrt{1-\left(\frac{\omega_j}{2}\right)^2}\
\beta(\rho_j,\rho_{j+1})\right]
\end{eqnarray}
for $|\omega_j| < 2$ and
\begin{eqnarray}
\label{deromegafinal:2}
\frac{\partial S_{\rm cl}}{\partial\omega_j}
& = &
\frac{1}{\pi\sqrt{\omega_j^2-4}}\
{\arcsin}\left[\sqrt{\left(\frac{\omega_j}{2}\right)^2-1}\
\beta(\rho_j,\rho_{j+1})\right]
\end{eqnarray}
for $|\omega_j| > 2$, where
$$
\beta(z,w)\stackrel{\rm def}{=}
\frac{|z - w|}{2\sqrt{\Im\,z\Im\,w}}\,.
$$

\section{Polyakov conjecture}
\setcounter{equation}{0}

The Polyakov conjecture states that
\begin{eqnarray}
\label{der_z}
\frac{\partial S_{\rm cl}}{\partial z_i}
& = & - c_i,
\hskip 5mm
i=1,\dots,m;\\
\label{der_x}
\frac{\partial S_{\rm cl}}{\partial x_j}
& = & -c^{\rm \scriptscriptstyle B}_j,
\hskip 5mm
j=1,\dots,n.
\end{eqnarray}
The equations (\ref{der_z}) can be derived by essentially the same methods one
uses in proving the Polyakov conjecture for the Riemann sphere
and we shall skip here the derivation. Let us only mention that the
equations (\ref{der_z}) are valid
in the case of parabolic and hyperbolic bulk singularities as well, although the classical
Liouville action is different in those cases.

We shall prove the equations (\ref{der_x}). Using the Liouville equation (\ref{bulk}) and the boundary conditions
(\ref{boundary1}) one gets
\begin{eqnarray}
\label{derivative}
\frac{\partial S_{\rm cl}}{\partial x_j}
 & = &
\lim_{\epsilon\to 0} D_{\epsilon}[\varphi],
\nonumber
\\
\nonumber
D_{\epsilon}[\varphi]
& = &
\frac{i}{8\pi}\int\limits_{s_j}
\left(\bar\partial\varphi\,d\bar z  - \partial\varphi\,dz\right)
\frac{\partial\varphi}{\partial x_j}
+
\frac{1-\nu_j}{4\pi}\int\limits_{s_j}\kappa_z|dz|\
\frac{\partial\varphi}{\partial x_j}
\\
& + &
\frac{i}{8\pi}\int\limits_{s_j}(d\bar z - dz)
\left(
 \partial\varphi\bar\partial\varphi+
 m{\rm e}^{\varphi}\right)
+
\frac{1-\nu_j}{4\pi}\int\limits_{s_j}\!\kappa_z|dz|\ \partial_x\varphi \\
&+ &
\nonumber
\frac{m_{k-1}}{4\pi}\left.{\rm e}^{\varphi\over2}\right|_{z = x_j-\epsilon}
-
\frac{m_{k}}{4\pi}\;\left.{\rm e}^{\varphi\over 2}\right|_{z = x_j+\epsilon}.
\end{eqnarray}
Applying the identity
\[
\frac{\partial\varphi}{\partial x_j} = - \partial_x\varphi + h_j,
\]
where
$$
h_j= - \sum\limits_{k\neq j}\frac{\partial\varphi}{\partial x_k}
 - \sum\limits_{i=1}^m\left(\frac{\partial\varphi}{\partial z_i} +
\frac{\partial\varphi}{\partial \bar z_i} \right),
$$
one has
\[
\frac{1-\nu_j}{4\pi}\int\limits_{s_j}\kappa_z|dz|\
\frac{\partial\varphi}{\partial x_j}
+
\frac{1-\nu_j}{4\pi}\int\limits_{s_j}\!\kappa_z|dz|\
\partial_x\varphi
=
\frac{1-\nu_j}{4\pi}\int\limits_{s_j}\!\kappa_z|dz|\
h_j,
\]
and
\begin{eqnarray*}
\frac{i}{8\pi}\int\limits_{s_j}
\left(\bar\partial\varphi\,d\bar z  - \partial\varphi\,dz\right)
\frac{\partial\varphi}{\partial x_j}
& = &
 \\
& &\hspace{-130pt} =\;
-\frac{i}{8\pi}\int\limits_{s_j}
\left(\bar\partial\varphi\,d\bar z  - \partial\varphi\,dz\right)
\left(\partial\varphi + \bar\partial\varphi\right)
 +
 \frac{i}{8\pi}\int\limits_{s_j}
\left(\bar\partial\varphi\,d\bar z  - \partial\varphi\,dz\right)
h_j.
\end{eqnarray*}
Note that the function $h_j$ is regular for  $z\to x_j$.
Taking into account the asymptotic behavior of $\varphi$ for $z\to x_j$
one thus gets in the limit $\epsilon \to 0$:
$$
\frac{1-\nu_j}{4\pi}\int\limits_{s_j}\!\kappa_z|dz|\
h_j +
\frac{i}{8\pi}\int\limits_{s_j}
\left(\bar\partial\varphi\,d\bar z  - \partial\varphi\,dz\right)
h_j
= o(1).
$$
It follows that up to terms vanishing in the limit $\epsilon \to 0$
the first two lines on the r.h.s.\ of (\ref{derivative}) yield:
\begin{eqnarray*}
&&
\hspace*{-2cm}
-\frac{i}{8\pi}\int\limits_{s_j}
\left(\bar\partial\varphi\,d\bar z  - \partial\varphi\,dz\right)
\left(\partial\varphi + \bar\partial\varphi\right)
+
\frac{i}{8\pi}\int\limits_{s_j}(d\bar z - dz)
\left(
 \partial\varphi\bar\partial\varphi +
 m{\rm e}^{\varphi}\right) \\
&= &
\frac{i}{8\pi}\int\limits_{s_j}\!dz
\left((\partial\varphi)^2 - m{\rm e}^{\varphi}\right)
-
\frac{i}{8\pi}\int\limits_{s_j}\!d\bar z
\left((\bar\partial\varphi)^2 - m{\rm e}^{\varphi}\right)
 \\
&= &
\frac{i}{4\pi}\int\limits_{s_j}\!dz
\left(\partial^2\varphi - \partial\bar\partial\varphi
- 2T_{\rm cl}(z)\right)
-
\frac{i}{4\pi}\int\limits_{s_j}\!d\bar z
\left(\bar\partial^2\varphi - \partial\bar\partial\varphi
- 2\overline{T}_{\rm cl}(\bar z)\right),
\end{eqnarray*}
where we have used the expression (\ref{Tclass})
for the classical energy-momentum tensor
and its complex conjugate $\overline{T}_{\rm cl}(\bar z)$ along with the Liouville equation (\ref{bulk}).
Since
$
\overline{T}_{\rm cl}(\bar z) \; = \; T_{\rm cl}(\bar z)
$
we have
\[
-  \frac{i}{2\pi}\int\limits_{s_j}\!dz\  T_{\rm cl}(z)
+  \frac{i}{2\pi}\int\limits_{s_j}\!d\bar z\
\overline{T}_{\rm cl}(\bar z)
\; = \;
 \frac{i}{2\pi}\!\!\!\!\!\!\int\limits_{|z-x_j| = \epsilon}\!\!\!\!\!\!
   dz\ T_{\rm cl}(z)
\; = \;
- c^{\rm \scriptscriptstyle B}_j .
\]
On the other hand
\begin{eqnarray*}
&&
\hspace*{-1.5cm}
\frac{i}{4\pi}\int\limits_{s_j}\!dz
\left(\partial^2\varphi - \partial\bar\partial\varphi
\right)
-
\frac{i}{4\pi}\int\limits_{s_j}\!d\bar z
\left(\bar\partial^2\varphi - \partial\bar\partial\varphi
\right)\\
&=&
\frac{i}{4\pi}\int\limits_{s_j}
(dz\partial + d\bar z\bar\partial)
\left(\partial\varphi - \bar\partial\varphi\right)
\; = \;
\frac{1}{4\pi}\int\limits_{s_j}\!d\vartheta\
\frac{\partial}{\partial\vartheta}\ \partial_y\varphi\\
&=&
\frac{1}{4\pi}\Big(
\partial_y\varphi|_{x = x_j+\epsilon}
-
\partial_y\varphi|_{x = x_j-\epsilon}
\Big)
\;=\;
\frac{m_{k}}{4\pi}\;\left.{\rm e}^{\varphi\over 2}\right|_{z = x_j+\epsilon}
-
\frac{m_{k-1}}{4\pi}\left.{\rm e}^{\varphi\over2}\right|_{z = x_j-\epsilon},
\end{eqnarray*}
so that one finally gets
\[
D_{\epsilon}[\varphi] \; = \; -c^{\rm \scriptscriptstyle B}_j + o(1),
\]
and
\[
\frac{\partial S_{\rm cl}}{\partial x_j}
\; = \;
-c^{\rm \scriptscriptstyle B}_j.
\]
Let us note that the proof presented  above applies to the bulk singularities as well.

\section{Classical solutions}
\setcounter{equation}{0}

\subsection{One bulk elliptic singularity}

In the case  of an elliptic singularity located at $z_1$ on the upper half-plane
$\Im z_1>0$ the energy-momentum takes the
 form
\begin{equation}
\label{T:bulk:elliptic} T(z) = \frac{1-\xi^2}{4} \left(
\frac{1}{(z-z_1)^2} + \frac{1}{\left(z-{\bar z_1}\right)^2}
-\frac{2}{(z-z_1)\left(z-\bar z_1\right)} \right),
\hskip 5mm
0<\xi <1.
\end{equation}
Since the singularity is elliptic and there are no singularities on the real axis
it is more convenient to work with the normalized solutions to the Fuchs equation
(\ref{Fuchs})
with a diagonal $SU(1,1)$ monodromy:
\begin{equation}
\label{basis:one:elliptic}
\Psi
=
\left[\begin{array}{r}
\psi_1 \\
\psi_2
\end{array}
\right]
=
\left[\begin{array}{r}
\frac{1}{\sqrt{\xi\left(z_1-\bar z_1\right)}}
(z-z_1)^{\frac{1-\xi}{2}}
(z-\bar z_1)^{\frac{1+\xi}{2}} \\
\frac{1}{\sqrt{\xi\left(z_1-\bar z_1\right)}}
(z-z_1)^{\frac{1+\xi}{2}}
(z-\bar z_1)^{\frac{1-\xi}{2}}
\end{array}
\right].
\end{equation}
In this setting a real, single valued conformal factor of the hyperbolic geometry
with constant scalar curvature $-2m$
is given by
\begin{eqnarray}
\label{metric:e:1}
{\rm e}^{-\varphi(z,\bar z)/2}
& = &
\frac{\sqrt{m}}{2}
\left(
{\rm e}^{t}\psi_1( z)\overline{\psi_1(z)} - {\rm e}^{-t}\psi_2(z)\overline{\psi_2(z)}
\right) \\
& = &
\frac{\sqrt{m}}{2\xi}
\left|\frac{(z-z_1)(z-\bar z_1)}{z_1-\bar z_1}\right|
\left[
{\rm e}^{t}\left|\frac{z-z_1}{z-\bar z_1}\right|^{-\xi}
-
{\rm e}^{-t}\left|\frac{z-z_1}{z-\bar z_1}\right|^{\xi}
\right],\nonumber
\end{eqnarray}
where $t$ is real and $t\geqslant 0$ so that the r.h.s.\ is non negative on the upper half plane
and the real axis.
By direct calculations one gets
$$
\left.\partial_y {\rm e}^{-{\varphi\over 2}}\right|_{\mathbb{R}}
={\sqrt{m}\over 2}\left( {\rm e}^{t} +{\rm e}^{-t}\right)=\sqrt{m}\cosh t = -{m_{\rm \scriptscriptstyle B}\over 2}\,.
$$
Thus a regular metric can be constructed if and only if
 the bulk and boundary cosmological constants  satisfy the conditions
$$
m_{\rm \scriptscriptstyle B}<0,
\hskip 5mm
m_{\rm \scriptscriptstyle B}^2\geqslant 4 m.
$$
This  in particular means that the geodesic curvature $\kappa$ of the boundary
is bounded from below,
$
\kappa\geqslant m.
$

For the solution (\ref{metric:e:1}) the r.h.s. of the equations
(\ref{der_xi}), (\ref{der_omega}), (\ref{der_z}), and (\ref{der_m}) can be easily calculated,
\begin{eqnarray*}
\frac{\partial S_{\rm cl}}{\partial\xi}  &=&
\log{\xi}
-\log{\textstyle \frac{\sqrt{m}}{2}}
-t
-\xi \log \left|z_1-\bar z_1\right|,
\\
\frac{\partial S_{\rm cl}}{\partial\omega_{\rm \scriptscriptstyle B}}
&=&
{\xi\over \sqrt{{\omega_{\rm \scriptscriptstyle B}^2} -4}},
\\
\frac{\partial S_{\rm cl}}{\partial z_1}&=&{1-\xi^2\over 2}\frac{1}{z_1-\bar z_1},
\end{eqnarray*}
where $\omega_{\rm \scriptscriptstyle B} = - 2 \cosh t$.
The solution reads\footnote{In the case under consideration
the classical action could be also calculated by explicit integration in (\ref{action}) (see \cite{Menotti:2006tc}).}
\begin{equation}
\label{scl_e_1}
S_{\rm cl} =
\xi\log{\xi}
- \xi
-{\xi \over 2}\log m
+\xi \log 2
-\xi t
-{1- \xi^2\over 2} \log \left|z_1-\bar z_1\right|
+
{\rm const}.
\end{equation}
Let us note in passing that using the form of the elliptic basis (\ref{basis:one:elliptic}) one gets for the map
(\ref{map:rho})
\[
\rho(z) = i\frac{{\rm e}^{\frac{t}{2}}\psi_1(z) - {\rm e}^{-\frac{t}{2}}\psi_2(z)}
{{\rm e}^{\frac{t}{2}}\psi_1(z) - {\rm e}^{-\frac{t}{2}}\psi_2(z)}
=
i\frac{{\rm e}^{t}(z-\bar z_1)^\xi - (z-z_1)^\xi}{{\rm e}^{t}(z-\bar z_1)^\xi - (z-z_1)^\xi}\,,
\]
so that
\[
\rho_1 = \lim_{z\to -\infty}\rho(z) = i \frac{\sinh t + i \sin 2\pi\xi}{\cosh t + \cos 2\pi\xi}\,,
\hskip 1cm
\rho_2 = \lim_{z\to \infty}\rho(z) = i \frac{\sinh t}{\cosh t +1}\,.
\]
This immediately gives
\[
\frac{|\rho_{j+1} - \rho_j|}{2\sqrt{\Im\,\rho_{j+1}\Im\,\rho_j}}
\; = \;
\frac{\sin\pi\xi}{|\sinh t|}
\; = \;
\frac{\sin\pi\xi}{\sqrt{\left(\frac{\omega_{\rm \scriptscriptstyle B}}{2}\right)^2-1}}
\]
and from (\ref{deromegafinal:2}) we get as above
\[
\frac{\partial S_{\rm cl}}{\partial\omega_{\rm \scriptscriptstyle B}}
\; = \;
{\xi\over \sqrt{{\omega_{\rm \scriptscriptstyle B}^2} -4}}\,,
\]
which confirms formulae (\ref{deromegafinal:1}) and
(\ref{deromegafinal:2}).

We now shall compare (\ref{scl_e_1}) with the semi-classical limit of the
FZZ 1-point function:
\begin{eqnarray}
\label{DOZZ:1point}
\nonumber
\langle\,V_\alpha(z_1)\,\rangle&=&
{U(\alpha,\mu_{\rm \scriptscriptstyle B})\over\left|z_1-\bar z_1\right|^{2\Delta_\alpha}}\,,\\
U(\alpha,\mu_{\rm \scriptscriptstyle B}) &=& {2\over b} (\pi \mu \gamma(b^2))^{Q-2\alpha\over 2b}
\Gamma(2b\alpha - b^2) \Gamma \left({2\alpha\over b} -{1\over b^2} -1\right)
\cosh(2\alpha -Q)\pi s,
\end{eqnarray}
where $s$ is defined by the relation
\begin{equation}
\label{def_s}
\cosh^2 \pi b s = {\mu_{\rm \scriptscriptstyle B}^2 \over \mu } \sin \pi b^2.
\end{equation}
In our notation
\begin{eqnarray}
\label{our:notation}
\mu&=&{m\over 4\pi b^2}\,,
\hskip 5mm
\mu_{\rm \scriptscriptstyle B}={m_{\rm \scriptscriptstyle B}\over 4\pi b^2 }\,,
\hskip 5mm
\alpha= {Q\over 2}(1-\xi),
\hskip 5mm
Q=b+{1\over b}\,,
\end{eqnarray}
and
\begin{eqnarray*}
U(\xi,m_{\rm \scriptscriptstyle B}) &=&
{2\over b} (\pi \mu \gamma(b^2))^{{1\over 2}(1+{1\over b^2})\xi}
\Gamma(b^2(1-\xi) +1-\xi) \Gamma \left(-(1+\textstyle{1\over b^2})\xi \right)
\cosh(-(b+\textstyle{1\over b})\xi \pi s).
\end{eqnarray*}
In  the limit $b \to 0$ the relation (\ref{def_s}) yields
$
\pi b s= t,
$
hence
\[
{2\over b} (\pi \mu \gamma(b^2))^{{1\over 2}(1+{1\over b^2})\xi}
\Gamma(b^2(1-\xi) +1-\xi)
\cosh(-(b+\textstyle{1\over b})\xi \pi s)
\approx
{\rm e}^{-{1\over b^2}
\left[
- \xi
+\xi \log \xi
-
\xi \log b^2-t\xi
\right]}.
\]
Due to the poles of the gamma function along the negative real axis
the $b\to 0$ asymptotic of the term
$
\Gamma \left(-(1+\textstyle{1\over b^2})\xi \right)
$
is more subtle. Within the path integral approach
the poles in (\ref{DOZZ:1point}) arise due to the integration over the zero mode of the Liouville field.
If the classical solution exists,  one should not
expect any pole structure in the quasi-classical limit.
In order to show that the poles can be regarded as
a sub-leading correction one may use the formula
$$
\Gamma(-x) =- {\pi\over x \Gamma(x) \sin \pi x}\,,
$$
along with the Stirling asymptotic expansion
$$
x\Gamma (x) \approx {\rm e}^{-x + \left(x+\textstyle{1\over 2}\right) \log x }.
$$
In the limit $b\to 0$ one obtains:
\begin{eqnarray*}
\Gamma \left(-(1+\textstyle{1\over b^2})\xi \right)
&\approx&
{\rm e}^{{1\over b^2}\xi
-\left({1\over b^2}\xi+{1\over 2} \right)\log {1\over b^2}\xi
-\log \left(-  {1\over \pi }\sin {\pi \over b^2}\xi\right)}\\
&=&
{\rm e}^{-{1\over b^2}\left[
- \xi
+\xi \log \xi
-\xi \log  b^2
+{1\over 2}b^2 \log {1\over b^2}\xi
+b^2\log \left(-  {1\over \pi }\sin {\pi \over b^2}\xi\right)\right]}.
\end{eqnarray*}
Keeping only the leading terms one thus have
$$
U(\xi,m_{\rm \scriptscriptstyle B})
\approx
{\rm e}^{-{1\over b^2} (
-{\xi\over 2} \log m +\xi \log 2
- \xi
+\xi \log \xi- t\xi)},
$$
in perfect agreement with the classical action (\ref{scl_e_1}).

\subsection{One bulk hyperbolic singularity}

Hyperbolic weight corresponds to the energy-momentum of the form
\begin{equation}
\label{T:bulk:hip}
T(z) = \frac{1+\lambda^2}{4}
\left(
\frac{1}{(z-z_1)^2} + \frac{1}{(z-\bar z_1)^2}
-\frac{2}{(z-z_1)(z-\bar z_1)}
\right).
\end{equation}
Repeating (with obvious modifications) the calculations from the
previous subsection one gets the metric
\begin{equation}
\label{metric:h:2}
{\rm e}^{-\varphi(z,\bar z)/2}
\; = \;
\frac{\sqrt{m}}{\lambda}
\left|\frac{(z-z_1)(z-\bar z_1)}{z_1-\bar z_1}\right|
\sin\left[
\lambda\log\left|\frac{z- z_1}{z-\bar z_1}\right| - t
\right],
\end{equation}
where
\begin{equation}
\label{s} \cos t = -\frac{\omega_{\rm \scriptscriptstyle B}}{2}
\equiv -\frac{m_{\rm \scriptscriptstyle B}}{2\sqrt{m}}\,.
\end{equation}
The metric can be constructed only if \( |\omega_b| \leq 2\).
In the coordinates
\[
w = \tau + i\sigma = \log \frac{z- z_1}{z-\bar z_1}
\]
it takes the form
\begin{equation}
\label{metric:h:3} {\rm e}^{-\varphi(w,\bar w)/2} \; = \;
\frac{\sqrt{m}}{\lambda}\sin(\lambda\tau - t).
\end{equation}
The conformal factor
is singular along the lines
\[
\lambda \tau = t + \pi k,
\hskip .5cm k \in {\mathbb Z},
\]
and the metric ${\rm e}^{\varphi(w,\bar w)}|dw|^2$ has closed geodesics located at
\[
\lambda\tau = t + \frac{\pi}{2}(2k +1),
\hskip .5cm k \in {\mathbb Z}.
\]
For positive $\omega_b$  there exists a solution of (\ref{s}) satisfying
\(
-\pi  <  t  <  -\frac{\pi}{2}.
\)
The metric between the real axis $\tau = 0$ and the geodesic
corresponding to $k=0,$
\[
\lambda\tau_g = \frac{\pi}{2}+ t  < 0,
\]
is then regular. As a final step of the construction of the ${\cal C}^1$ metric on
the upper half-plane we shall ``fill in'' the hole $\tau < \tau_g$ with a
flat metric determined by
\[
{\rm e}^{-\varphi_0(w,\bar w)/2}
\; = \;
\frac{\sqrt{m}}{\lambda}
\]
or, in the $z$ coordinates,
\[
{\rm e}^{-\varphi_0(z,\bar z)/2}
\; = \;
\frac{\sqrt{m}}{\lambda}
\left|\frac{(z- z_1)(z-\bar z_1)}{ z_1-\bar z_1}\right|.
\]

The classical Liouville action in the presence of hyperbolic singularities
\cite{Hadasz:2003kp,Hadasz:2003he}
is constructed as a sum of the standard Liouville action functional calculated on the conformal factor of
the hyperbolic metric in the region ``between the holes''
and the actions for ``holes'' around each hyperbolic singularity. In our case the
first contribution is
\begin{eqnarray}
\label{action:outside}
\hspace*{-5mm}
S_1(m,m_{\rm \scriptscriptstyle B},\lambda, z_1)
& \! = \! & \frac{1}{4\pi}\lim_{R\to \infty}\left[
\int\limits_{{\cal M}_R}\!\!\!d^2z
\left[|\partial\varphi|^2 + m{\rm e}^{\varphi}\right]
+
m_{\rm \scriptscriptstyle B}\int\limits_{-R}^R\!\!dy\; {\rm e}^{\varphi/2}
+
\int\limits_{\partial{\cal M}_R}\!\!\!\!\kappa_z|dz|\;\varphi\right]\!,
\;\;\;\;\;\;
\end{eqnarray}
where ${\cal M}_R$ is a part of the upper half plane outside the hole,
\(
\log\left|\frac{z-z_1}{z-\bar z_1}\right| > \frac{1}{\lambda}\left(\frac{\pi}{2}+ t\right),
\)
bounded by the semi-circle of radius $R.$
The second one is a regularized Liouville action functional, calculated for the flat metric $\varphi_0$ on
the hole around $z = z_1$ with a small disc of radius $\epsilon$  removed,
\[
H_\epsilon = \left\{z\, \in\,{\mathbb H}:\;\;
\log\left|\frac{z-z_1}{z-\bar z_1}\right| < \frac{1}{\lambda}\left(\frac{\pi}{2}+ t\right)
\;\; \wedge\;\;  |z - z_1| < \epsilon
\right\}.
\]
It reads
\begin{eqnarray}
\label{action:hole}
S_2(m,m_{\rm \scriptscriptstyle B},\lambda, z_1)
& = &
\lim_{\epsilon\to 0}
S_{2,\epsilon}(m,m_{\rm \scriptscriptstyle B},\lambda, z_1), \\
S_{2,\epsilon}(m,m_{\rm \scriptscriptstyle B},\lambda, z_1)
& = &
\frac{1}{4\pi}\int\limits_{H_\epsilon}\!d^2z\;
\left[|\partial\varphi_0|^2 + m{\rm e}^{\varphi_0}\right]
+
\frac{1}{4\pi}\int\limits_{\partial H_\epsilon} \kappa_z|dz|\varphi_0
+(\lambda^2 -1)\log\epsilon.
\nonumber
\end{eqnarray}
Shifting
\[
\varphi = \tilde\varphi - \log m,
\hskip 1cm
\varphi_0 = \tilde\varphi_0 - \log m,
\]
one checks  that the classical action depends on $m$ only trough
$\omega_{\rm \scriptscriptstyle B},$
\[
S_{\rm cl}
\; = \;
S\left(\frac{m_{\rm \scriptscriptstyle B}}{\sqrt m},\lambda, z_1\right)
\; \equiv \;
S\left(\omega_{\rm \scriptscriptstyle B},\lambda, z_1\right).
\]
From the Polyakov conjecture and eq.\ (\ref{T:bulk:hip}) one gets
\begin{equation}
\label{der1}
\frac{\partial S_{\rm cl}}{\partial z_1}
\; = \;
\frac{1+\lambda^2}{2}\frac{1}{z_1 - \bar z_1}\,.
\end{equation}
One also has (using the form of functionals
(\ref{action:outside}) and (\ref{action:hole}))
\begin{eqnarray}
\label{der2}
\frac{\partial S_{\rm cl}}{\partial \omega_{\rm \scriptscriptstyle B}}
& = &
\frac{1}{4\pi}\int\limits_{\mathbb R}\!dy\; {\rm e}^{\tilde\varphi/2}
\; = \;
\frac{\lambda}{\sqrt{4-\omega_{\rm \scriptscriptstyle B}^2}}\,,
\end{eqnarray}
and finally
\begin{eqnarray}
\label{der3}
\frac{\partial S_{\rm cl}}{\partial\lambda}
& = &
\lim_{\epsilon\to 0}
\left[
\frac{1}{4\pi}\int\limits_{H_\epsilon}\!d^2z\;
\frac{\partial}{\partial\lambda}{\rm e}^{\tilde\varphi_0}
\; + \; \lambda\log\epsilon\right] \\
\nonumber
& = &
\lim_{\epsilon\to 0}
\left[
\int\limits_{\tau_0}^{\tau g}\!d\tau\;
\frac12\frac{\partial}{\partial\lambda} \lambda^2 \; + \; \lambda\log\epsilon\right]
\; = \;
\left(t+\frac{\pi}{2}\right) + \lambda\log| z_1 - \bar z_1|,
\end{eqnarray}
where the change of the integration variables from $z$ to $w$ and the fact that
\(
|z-\xi| = \epsilon
\)
corresponds to
\(
\tau \; = \;  \log\epsilon - \log| z_1 - \bar z_1|
\)
have been used.

Integrating the equations (\ref{der1}), (\ref{der2}) and (\ref{der3}) we get
\begin{equation}
\label{class:action:final}
S(m,m_{\rm \scriptscriptstyle B},\lambda, z_1)
\; = \;
\frac{1+\lambda^2}{4}\log| z_1 - \bar z_1|^2
+\lambda\left(t+\frac{\pi}{2}\right)
+{\rm const},
\end{equation}
where the constant is independent of $m,m_{\rm \scriptscriptstyle B},\lambda, z_1$.

For hyperbolic weights $\Delta = \alpha(Q-\alpha) > \frac{Q^2}{4},$ i.e.\ for $\alpha$  of the form
\[
\alpha = \frac{Q}{2}\left(1+i\lambda\right),
\hskip 1cm
\lambda\in {\mathbb R},
\]
the Liouville one-point coupling constant $U(\alpha)$ given by (\ref{DOZZ:1point}) is complex. Let us write
\[
U(\alpha) = \Phi(\alpha)U_{\rm s}(\alpha)
\]
where $\Phi$ is the phase and $U_{\rm s}$ the modulus\footnote{The subscript s is meant to remind that
$U_s(\alpha)$ is symmetric under reflection $\alpha \to
Q-\alpha$} of
$U.$ The phase $\Phi$ coincides with a square root of the
Liouville reflection amplitude,
\begin{eqnarray*}
\Phi^2(\alpha)
& = &
S_{\rm L}(\alpha),
\\
S_{\rm L}(\alpha)
& = &
\left(\pi\mu\gamma\left(b^2\right)\right)^{\frac{2\alpha -Q}{b}}
\frac{
\Gamma(-(2\alpha -Q)/b)\Gamma(1-b(2\alpha -Q))}
{\Gamma((2\alpha -Q)/b)\Gamma(1+b(2\alpha -Q))},
\end{eqnarray*}
and
\[
U_{\rm s}(\alpha)  =
\frac{2\cosh\left[(2\alpha - Q)\pi \sigma\right]}{b}\sqrt{
    \Gamma\!\left(1+(2\alpha -Q)b\right)
    \Gamma\!\left(1-(2\alpha -Q)b\right)
    \Gamma\!\left(\frac{2\alpha -Q}{b}\right)
    \Gamma\!\left(\frac{Q-2\alpha}{b}\right)}.
\]
Using (\ref{def_s}), (\ref{our:notation}) and the fact that $t < 0,$ we get for $b \to 0$
\[
\frac{\!\!U_{\rm s}(\alpha)}{| z_1-\bar z_1|^{2\Delta_{\alpha}}} \; \sim \;
\exp\left\{-\frac{1}{b^2} \left(
\frac{1+\lambda^2}{4}\log| z_1 - \bar z_1|^2
+\lambda\left(t+\frac{\pi}{2}\right)\right)\right\},
\]
in perfect agreement with the classical action (\ref{class:action:final}) again.
\subsection{Two boundary elliptic singularities}

In the case of two singularities the conformal weights
must be the same, $\nu_1=\nu_2=\nu$. Using the $SL(2,\mathbb{R})$
symmetry one can place them at $x_1=0$ and $x_2=\infty$.
This corresponds to the following energy-momentum tensor
$$
T(z)={1-\nu^2\over 4}{1\over z^2}\,.
$$
Normalized solutions, regular and real along the positive and the  negative semi-axes,
are given by
\begin{equation}
\label{solpositive}
\Psi_1 =
\left[
\begin{array}{c}
\psi_1^-(z)\\
\psi_1^+(z)
\end{array}
\right]
=
\left[
\begin{array}{c}
\frac{1}{\sqrt{\nu}}z^{\frac{1-\nu}{2}}\\
 \frac{1}{\sqrt{\nu}}z^{\frac{1+\nu}{2}}
\end{array}
\right],\;\;\;\;
\Psi_2 =
\left[
\begin{array}{c}
\psi_2^-(z)\\
\psi_2^+(z)
\end{array}
\right]
=
\left[
\begin{array}{c}
\frac{1}{\sqrt{\nu}}
 (-z)^{\frac{1+\nu}{2}}\\
\frac{1}{\sqrt{\nu}}
(-z)^{\frac{1-\nu}{2}}
\end{array}
\right],
\end{equation}
respectively. They are related on the upper half plane
by
\begin{equation}
\label{rel_1_2}
\Psi_1=
\left[
\begin{array}{cc}
0&{\rm e}^{{i\pi (1-\nu)\over 2}}\\
{\rm e}^{{i\pi (1+\nu)\over 2}}&0
\end{array}
\right]
 \Psi_2.
\end{equation}
In terms of $\Psi_1$ the solution to the Liouville equation reads
\begin{equation}
\label{sol_2boundary}
{\rm e}^{-\varphi/2} = \frac{\sqrt m}{2i} \Psi_1^{\rm T}(z) \Sigma M
\overline{\Psi_1( z)}\,,
\end{equation}
where the matrix
$$
M=\left[
\begin{array}{cc}
a& i \beta\\
i\gamma &\overline{a}
\end{array}
\right], \;\;\;|a|^2 +\beta\gamma = 1,\;\;\;\gamma,\beta\in \mathbb{R},\;a\in \mathbb{C},
$$
can be chosen such that the r.h.s.\ of (\ref{sol_2boundary})
is positive on the upper half plane. This implies in particular that $\gamma$ and $-\beta$ are
positive. The boundary conditions
\begin{equation}
\label{boundarybb1}
\left.\partial_y\varphi\right|_{y = 0} = m_1{\rm e}^{\varphi/2} \;\;\;{\rm for}\;\;\;
x>0,\;\;\;\;\;\;\;\;\;
\left.\partial_y\varphi\right|_{y = 0} = m_2{\rm e}^{\varphi/2} \;\;\;{\rm for}\;\;\;
x<0,
\end{equation}
imply
\begin{eqnarray*}
a  +\bar a &=& {m_1\over  \sqrt{m}}=\omega_1,\;\;\;\;\;\;\;\;\;
\alpha {\rm e}^{i\pi \nu}  +\bar \alpha{\rm e}^{-i\pi \nu} = -{m_2\over  \sqrt{m}}=-\omega_2.
\end{eqnarray*}
Solving for $a$ one gets
$$
a={-\omega_2 -{\rm e}^{-i\pi\nu} \omega_1\over 2 i \sin \pi \nu}.
$$

From the Gauss-Bonnet theorem one may expect that the geodesic curvature
of boundary components should be positive. It can be easily checked
that this is really so in the symmetric case $\omega_1=\omega_2=\omega,$
when the hyperbolic metric exists if the condition
$$
-\omega > 2 \sin {\textstyle{1\over 2}} \pi \nu
$$
is satisfied. We assume in the following  that the boundary geodesic curvatures are positive
and such that the function
\begin{equation}
\label{solplus}
{\rm e}^{-\varphi/2}
=
\frac{\sqrt{m}}{2\nu}|z|
\Big[
\gamma |z|^{-\nu} -\beta |z|^{\nu} -2\Im a {\rm e}^{i\nu \theta}
\Big]
\end{equation}
is positive for $z= |z|{\rm e}^{i\theta}$ in the upper half plane.
Introducing parametrization
\[
\omega_1 = -2 \cosh t_1,
\hskip 5mm
\omega_2 =-2 \cosh t_2,
\hskip 5mm
t_1,t_2\geqslant 0,
\]
we have for the solution
(\ref{solplus})
\begin{eqnarray*}
\frac{\partial S_{\rm cl}}{\partial\nu} & = &
\lim_{\epsilon\to 0}\left[
 -\frac{1}{4\pi}\hspace{-10pt}
 \int\limits_{|z|=\epsilon, \Im z>0}\hspace{-10pt}
 \kappa\,|dz|\, \varphi
  +{1-\nu\over 2}\log\epsilon
+\frac{1}{4\pi}\hspace{-10pt}
\int\limits_{|z|={1\over \epsilon}, \Im z>0}\hspace{-10pt}
\kappa\,|dz|\, \varphi
 -{1+\nu\over 2}\log\epsilon
  \right] \\ &=&
\log{\nu}
-\log{\textstyle \frac{\sqrt{m}}{2}}
-{\textstyle{1\over 2}} \log (|a|^2-1) \\
&=&
\log{\nu}
-\log{\textstyle \sqrt{m}}
+\log \sin{\pi\nu}\\
&&
-{\textstyle{1\over 2}}
\log \sin\left({\pi(1-\nu) + i(t_1+t_2)\over 2}\right)
-{\textstyle{1\over 2}}
\log\sin\left({\pi(1-\nu) - i(t_1+t_2)\over 2}\right)
\\
&&
-{\textstyle{1\over 2}}
\log\sin\left({\pi(1-\nu) + i(t_1-t_2)\over 2}\right)
-{\textstyle{1\over 2}}
\log\sin\left({\pi(1-\nu) - i(t_1-t_2)\over 2}\right),
\\
\frac{\partial S_{\rm cl}}{\partial\omega_1}
&=&
{\sqrt{m}\over 4\pi} \int\limits_{0}^\infty dx\, {\rm e}^{ \varphi \over 2}
=
\frac{i}{2\pi}\log\left[\frac{
\sin\left({\pi(1-\nu) +i(t_1+t_2)\over 2}\right)
\sin\left({\pi(1-\nu) +i(t_1-t_2)\over 2}\right)}{
\sin\left({\pi(1-\nu) -i(t_1+t_2)\over 2}\right)
\sin\left( {\pi(1-\nu) -i(t_1-t_2)\over 2}\right)}
\right]
{\partial t_1\over \partial \omega_1},
\\
\frac{\partial S_{\rm cl}}{\partial\omega_2}
&=&
{\sqrt{m}\over 4\pi}\!\int\limits_{-\infty}^0\!dx\, {\rm e}^{ \varphi \over 2}
=
\frac{i}{2\pi}\log\left[\frac{
\sin\left({\pi(1-\nu) + i(t_1+t_2)\over 2}\right)
\sin\left({\pi(1-\nu) - i(t_1-t_2)\over 2}\right)}{
\sin\left({\pi(1-\nu) - i(t_1+t_2)\over 2}\right)
\sin\left({\pi(1-\nu) + i(t_1-t_2)\over 2}\right)}
\right]
{\partial t_2\over \partial \omega_2}.
\end{eqnarray*}
Integrating these equations one gets (up to a constant)
\begin{equation}
\label{sclassical:2boundary}
S_{\rm cl} =
\nu\left(\log\nu - \frac12\log m -1\right)
+ \mathsf{s}(\pi\nu)
+\sum\limits_{\tau_1 = \pm}\sum\limits_{\tau_2 = \pm}
\mathsf{s}\!\left(\frac{\pi(1-\nu) + i(\tau_1 t_1 + \tau_2 t_2)}{2}\right),
\end{equation}
where
\[
\mathsf{s}(x) \stackrel{\rm def}{=} \frac{1}{\pi}\int\limits_{\pi/2}^x\!dy\,\log\sin y.
\]

The quantum boundary two-point coupling constant has the form \cite{Fateev:2000ik}:
\begin{eqnarray}
\label{quantum:two:point}
d(\beta|s_1,s_2)
& = &
\frac{
\left[\pi\mu\gamma\left(b^2\right)b^{2-2b^2}\right]^{(Q-2\beta)/2b}
\Gamma_b(2\beta-Q)\Gamma_b^{-1}(Q - 2\beta)}{
S_b\left(\beta + i\frac{s_1+s_2}{2}\right)
S_b\left(\beta + i\frac{s_1-s_2}{2}\right)
S_b\left(\beta - i\frac{s_1-s_2}{2}\right)
S_b\left(\beta - i\frac{s_1+s_2}{2}\right)}\,,\;\;\;\;
\end{eqnarray}
with appropriate counterparts of  relations (\ref{def_s}),
(\ref{our:notation}) assumed.  Taking into account
\[
t_i = \pi b s_i, \hskip 5mm
\beta \sim \frac{1}{2b}(1+\nu),
\]
and the asymptotic behavior
\begin{equation}
\label{s_asym}
\log S_b(x) \; \sim \; \frac{1}{b^2}\,\mathsf{s}(\pi b x) +\frac{\log 2}{b^2}\left(x b -\frac12\right),
\end{equation}
one can check that
 the semiclassical asymptotic of (\ref{quantum:two:point})
is  given by  the classical action (\ref{sclassical:2boundary}) indeed.

\subsection{One bulk, one boundary  elliptic singularities}

For $T(z)$ having a single pole at the real axis and a single pole in
the interior of the upper half plane one can always choose the
``boundary'' pole to be located at $z = \infty.$
The second pole we shall take at $z = z_1,\ {\Im}z_1 > 0$.
The energy-momentum tensor takes the form
\begin{eqnarray}
\label{Tbub}
T(z) & = & \frac{\delta}{(z-z_1)^2} + \frac{\delta}{(z-\bar z_1)^2} +
\frac{\delta_b-2\delta}{(z-z_1)(z-\bar z_1)}
\end{eqnarray}
with elliptic weights
\[
\delta = \frac{1-\xi^2}{2},\hskip 1cm
\delta_b = \frac{1-\nu^2}{2}.
\]
A normalized basis in the space of solutions of (\ref{Fuchs}),
with diagonal SU(1,1) monodromy around $z= z_1,$ can be chosen in the form
\begin{equation}
\label{solutions}
\begin{array}{rcl}
\psi_1(z) & = &
\frac{1}{\sqrt{\xi(z_1-\bar z_1)}}
(z-z_1)^{\frac{1-\xi}{2}}(z-\bar z_1)^{\frac{1+\xi}{2}}
{}_2F_1\left(\frac{1-\nu}{2},\frac{1+\nu}{2},1-\xi,
\frac{z_1-z}{z_1 - \bar z_1}\right),
\\[10pt]
\psi_2(z) & = &
\frac{1}{\sqrt{\xi(z_1-\bar z_1)}}
(z-z_1)^{\frac{1+\xi}{2}}(z-\bar z_1)^{\frac{1-\xi}{2}}
{}_2F_1\left(\frac{1-\nu}{2},\frac{1+\nu}{2},1+\xi,
\frac{z_1-z}{z_1 - \bar z_1}\right).
\end{array}
\end{equation}
The functions $\psi_{1,2}(z)$ are analytic in the vicinity of the real axis
(the cuts between the branching points $z=z_1,\ z=\infty$ and $z=\bar z_1$ can be
chosen such that they do not intersect the real axis).
A real, single-valued around $z=z_1$ solution to the Liouville
equation (\ref{bulk}) can be
expressed through $\psi_1(z), \psi_2(z)$ as
\begin{equation}
\label{first:form}
{\rm e}^{-\varphi/2} =
\frac{\sqrt m}{2} \left(a|\psi_1(z)|^2 - a^{-1}|\psi_2(z)|^2\right),
\end{equation}
with a (real) constant $a$ to be determined from (\ref{boundary2}).

To  this end it is convenient  to express $\overline{\psi_i(z)}$
in terms of  $\psi_i(\bar z)$. Using the formulae for analytic
continuation of the hypergeometric functions one gets:
\begin{equation}
\label{cont}
 \left[
\begin{array}{cc}
\overline{\psi_1(z)} \\  \overline{\psi_2(z)}
\end{array}\right]
=
{\cal C}
\left[
\begin{array}{cc}
\psi_1(\bar z) \\ \psi_2(\bar z)
\end{array}\right]\!,
\end{equation}
where
\begin{equation}
{\cal C} =
i
\left[
\begin{array}{cc}
\frac{\Gamma(1-\xi)\Gamma(\xi)}
{\Gamma\left(\frac{1-\nu}{2}\right)\Gamma\left(\frac{1+\nu}{2}\right)}
&
\frac{\Gamma(1-\xi)\Gamma(-\xi)}
{\Gamma\left(\frac{1-\nu}{2}-\xi\right)\Gamma\left(\frac{1+\nu}{2}-\xi\right)}
\\
\frac{\Gamma(1+\xi)\Gamma(\xi)}
{\Gamma\left(\frac{1-\nu}{2}+\xi\right)\Gamma\left(\frac{1+\nu}{2}+\xi\right)}
&
\frac{\Gamma(1+\xi)\Gamma(-\xi)}
{\Gamma\left(\frac{1-\nu}{2}\right)\Gamma\left(\frac{1+\nu}{2}\right)}
\end{array}\right]\!.
\end{equation}
Hence
\begin{equation}
\label{second:form}
{\rm e}^{-\varphi/2}
=
\frac{\sqrt m}{2} \Psi(z)^{\rm T}\cdot A \cdot {\cal C} \cdot \Psi(\bar z),
\end{equation}
where $A = {\rm diag}\left(a,-a^{-1}\right)$.  The boundary condition
\[
\left.\partial_y{\rm e}^{-\varphi/2}\right|_{y=0}
\; = \; -\frac{m_{\rm \scriptscriptstyle B}}{2}
\; \equiv \; -\frac{\sqrt{m}}{2}\omega_{\rm \scriptscriptstyle B}
\]
yields
\begin{eqnarray}
\label{trace2}
\omega_{\rm \scriptscriptstyle B}
& = &
-a \frac{\Gamma(1-\xi)\Gamma(-\xi)}
{\Gamma\left(\frac{1-\nu}{2}-\xi\right)\Gamma\left(\frac{1+\nu}{2}-\xi\right)}
- a^{-1}\frac{\Gamma(1+\xi)\Gamma(\xi)}
{\Gamma\left(\frac{1-\nu}{2}+\xi\right)\Gamma\left(\frac{1+\nu}{2}+\xi\right)}\,.
\end{eqnarray}
Solving with respect to $a$ one gets
\begin{eqnarray}
\label{a}
a_\pm&=&
\frac{
\Gamma\left(\frac{1-\nu}{2}-\xi\right)
\Gamma\left(\frac{1+\nu}{2}-\xi\right)
\Gamma(1+\xi)
}
{\pi\Gamma(1-\xi)}\\
\nonumber
&\times &
\left(
{\omega_{\rm \scriptscriptstyle B}\over 2}
\sin\pi\xi
\pm\sqrt{
\cos^2\frac{\pi\nu}{2}
+\left({\omega_{\rm \scriptscriptstyle B}^2\over 4}-1
\right)\sin^2\pi\xi} \right).
\end{eqnarray}

Let us note that the change of sign in $a$ is equivalent to the change of sign
of $\omega_{\rm \scriptscriptstyle B}$ and the change of sing of the r.h.s.\ of
(\ref{first:form}). It does not lead therefore to any new solutions of the Liouville
equation. With no loss of generality we can then  work with  $a=a_+$.

It should be stressed that not for all parameters $\omega_{\rm \scriptscriptstyle B}, \xi,\nu,$
for which $a$ is real, the formula (\ref{first:form}) yields a regular solution
for the Liouville equation. Indeed, the r.h.s.\ of (\ref{first:form}) may change sign on
the upper half plane and the zero lines appearing in this situation give rise to singular lines of the corresponding
hyperbolic geometry. Even in the simple situation at hand, the problem of determining the
ranges of parameters for which a regular solution exists and its classical action is
well defined is rather involved and
we are not aware of any compete solution to it. In the following we simply assume that
$\omega_{\rm \scriptscriptstyle B}, \xi,\nu$ are such that a regular solution does exist.

Taking into account the asymptotic behavior of the solution $\Psi$ (\ref{solutions})
for $z\to z_1$ and (\ref{a})
one gets from (\ref{der_xi}):
\begin{eqnarray}
\nonumber
 \frac{\partial S_{\rm cl}}{\partial\xi} & = &
-{1\over 2}\log m
 - \xi\log|z_1-\bar z_1|
 +\log 2 \\
 \label{der:xi:bubo}
&-&
\log\left[ \frac{
\Gamma(\xi)
\Gamma\left(\frac{1+\nu}{2}-\xi\right)
}
{\Gamma(1-\xi)\Gamma\left(\frac{1+\nu}{2}+\xi\right)}
\right]
+ \log \cos\pi\left({\nu\over 2}+\xi\right)
\\
\nonumber
&- &
\log
\left[
{\omega_{\rm \scriptscriptstyle B}\over 2}
\sin\pi\xi
+\sqrt{
\cos^2\frac{\pi\nu}{2}
+\left({\omega_{\rm \scriptscriptstyle B}^2\over 4}-1
\right)\sin^2\pi\xi} \right].
\end{eqnarray}

Using the asymptotic behavior of the hypergeometric function for
large arguments one finds for $z \to \infty:$
\begin{eqnarray}
\label{asymptotes:of:psis}
\psi_1(z) & \sim &
\frac{\left(z_1-\bar z_1\right)^{-\frac{\nu}{2}}}{\sqrt\xi}
\frac{\Gamma(\nu)\Gamma(1-\xi)}{\Gamma\left(\frac{1+\nu}{2}\right)\Gamma\left(\frac{1+\nu}{2}-\xi\right)}\
(z - z_1)^{\frac{\nu - \xi}{2}} (z-\bar z_1)^{\frac{1+\xi}{2}},
\nonumber \\
\psi_2(z) & \sim &
\frac{(z_1-\bar z_1)^{-\frac{\nu}{2}}}{\sqrt\xi}
\frac{\Gamma(\nu)\Gamma(1+\xi)}{\Gamma\left(\frac{1+\nu}{2}\right)\Gamma\left(\frac{1+\nu}{2}+\xi\right)}\
(z - z_1)^{\frac{\nu + \xi}{2}} (z-\bar z_1)^{\frac{1-\xi}{2}},
\end{eqnarray}
and
\begin{eqnarray}
\label{der:nu:bubo}
\nonumber
\frac{\partial S_{\rm cl}}{\partial\nu}
& = & - \frac14\log m  +
\frac{\nu}{2}\log|z_1-\bar z_1| + \log\left[
\frac{\Gamma\left(\frac{1+\nu}{2}\right)}{\Gamma(\nu)}
\right]
-\frac12 \log\pi
\\
&+&
\textstyle\frac12 \log\left[ \Gamma\left(\frac{1+\nu}{2}-\xi\right)
 \Gamma\left(\frac{1+\nu}{2}+\xi\right)
 \right]+\frac12\log 2
 \\\nonumber
&-&
\frac12\log\left[
{B_+\over \sin\pi\xi \cos\pi\left({\nu\over 2}+\xi\right)}
-
{B_-\over \sin\pi\xi \cos\pi\left({\nu\over 2}-\xi\right)}
 \right],
\end{eqnarray}
where
\begin{eqnarray*}
B_\pm&=&
\pm{\omega_{\rm \scriptscriptstyle B}\over 2}
\sin\pi\xi
+
\sqrt{
\cos^2\frac{\pi\nu}{2}
+\left({\omega_{\rm \scriptscriptstyle B}^2\over 4}-1
\right)\sin^2\pi\xi}\ .
\end{eqnarray*}

In order to calculate the $\omega$-derivative one needs the map $\rho$
(\ref{map:rho})
defined in terms of
the $SL(2,\mathbb{R})$-monodromy solution to the Fuchs equation, which can be
easily constructed from the $SU(1,1)$-monodromy one:
$$
\tilde\Psi(z) =
\left[
\begin{array}{c}
\tilde\psi_1(z)\\
\tilde\psi_2(z)
\end{array}
\right]
=
{1\over \sqrt 2 i}
\left[
\begin{array}{rr}
i\sqrt{a} & -{i\over \sqrt{a}}\\
\sqrt{a} & {1\over \sqrt{a}}
\end{array}
\right]
\left[
\begin{array}{c}
\psi_1(z)\\
\psi_2(z)
\end{array}
\right].
$$
Using (\ref{asymptotes:of:psis}) one finds the $z\to \infty$ asymptotic
\[
\rho(z)
\; \equiv \;
{\tilde\psi_1(z)\over
\tilde\psi_2(z)}
\; \sim \;
i\ \frac{{\rm e}^{r} (z-z_1)^{-\xi}(z-\bar z_1)^\xi -1}{{\rm e}^{r} (z-z_1)^{-\xi}(z-\bar z_1)^\xi +1}\,,
\]
where
\begin{eqnarray*}
{\rm e}^r
& \equiv &
\frac{\Gamma(1-\xi)\Gamma\left(\frac{1+\nu}{2}+\xi\right)}
{\Gamma(1+\xi)\Gamma\left(\frac{1+\nu}{2}-\xi\right)}\, a \\
& = &
\frac{1}{\cos\left(\frac{\pi\nu}{2} + \pi\xi\right)}
\left[
\frac{\omega_{\rm \scriptscriptstyle B}}{2} \sin\pi\xi
+
\sqrt{
\cos^2\frac{\pi\nu}{2}
+\left({\omega_{\rm \scriptscriptstyle B}^2\over 4}-1
\right)\sin^2\pi\xi}
\right].
\end{eqnarray*}
Calculating the limits along the real axis
\[
\rho_1 = \lim_{x\to -\infty}\rho(x) = i\,\frac{\sinh r + i \sin 2\pi\xi}{\cosh r + \cos 2\pi\xi},
\hskip 1cm
\rho_2 = \lim_{x\to \infty}\rho(x) = i\, \frac{\sinh r}{\cosh r +1},
\]
one obtains
$$
\frac{|\rho_{j+1} - \rho_j|}{2\sqrt{\Im\,\rho_{j+1}\Im\,\rho_j}}
\; = \;
\frac{\sin\pi\xi}{\sinh r} .
$$
Hence for $\omega_{\rm \scriptscriptstyle B} > 2$
 formula (\ref{deromegafinal:2})  implies
\[
\frac{\partial S_{\rm cl}}{\partial\omega_{\rm \scriptscriptstyle B}}
\; = \;
\frac{1}{\pi\sqrt{\omega_{\rm \scriptscriptstyle B}^2-4}}\,
{\rm arcsin}\left[
\frac{\sin\pi\xi}{\sinh r}
\sqrt{\left(\frac{\omega_{\rm \scriptscriptstyle B}}{2}\right)^2-1}\right],
\]
or
\begin{equation}
\label{der:t:bubo}
\frac{\partial S_{\rm cl}}{\partial t}
\; = \;
\frac{1}{\pi}\,{\rm arcsin}\left[
\frac{\sinh t\sin\pi\xi}{\sinh r}\right],
\end{equation}
where $\omega_{\rm \scriptscriptstyle B} = 2\cosh t$.

Checking integrability conditions or direct integration of the equations
(\ref{der:xi:bubo}), (\ref{der:nu:bubo}) and (\ref{der:t:bubo})
is rather involved and not especially illuminating. We shall check instead that
(\ref{der:xi:bubo}), (\ref{der:nu:bubo}) and (\ref{der:t:bubo})
coincide with the corresponding derivatives of the classical
action obtained from the classical limit of the exact quantum expression.

The bulk-boundary correlation function
for the location of the bulk operator at $z=z_1$ and the
boundary operator at $z=\infty$ is given by \cite{Hosomichi:2001xc}:
\begin{eqnarray*}
\langle V_\alpha(z_1) B^{ss}_\beta (\infty)\rangle
&=&
{P(\alpha,\beta,|s)I(\alpha, \beta|s)
\over
|z_1 - \bar z_1|^{2\Delta_\alpha-\Delta_\beta}},
\\
P(\alpha, \beta|s)&=&
-2\pi i \left[
\pi\mu\gamma(b^2)b^{2-2b^2}
\right]^{{1\over 2b}(Q-2\alpha-\beta)}\\
&\times&
{\Gamma^3_b(Q-\beta)\Gamma_b(2\alpha -\beta)\Gamma_b(2Q-2\alpha-\beta)
\over
\Gamma_b(Q)\Gamma_b(Q-2\beta)\Gamma_b(\beta)\Gamma_b(Q-2\alpha)
\Gamma_b(2\alpha)
},
\\
I(\alpha, \beta|s) & = &
\int\limits_{i{\mathbb R}}\!du\; {\rm e}^{-2\pi u s}\
\frac{S_b(u + \beta/2 + \alpha - Q/2)S_b(u + \beta/2 - \alpha + Q/2)}
{S_b(u-\beta/2-\alpha+3Q/2)S_b(u-\beta/2 + \alpha + Q/2)},
\end{eqnarray*}
where  the relations (\ref{def_s}) and (\ref{our:notation}) are
still assumed.

Using the asymptotic of the Barnes gamma function
\[
\log \Gamma_b\left(\frac{x}{b}\right)
\sim -\frac{1}{b^2}\left\{
{\mathbf g}( x) + {1\over 2}\left(x - \frac12\right)\log 2\pi
+{1\over 2}\left(x-\frac12\right)^2 \log b
\right\},
\]
where
\[
{\mathbf g}(x) = \int\limits_{\frac{1}{2}}^x\! dy\ \log\Gamma (y),
\]
one obtains
\begin{eqnarray*}
\log P(\alpha, \beta|s)&\sim &
\textstyle{1\over b^2}\left[\textstyle
-{1\over 2}\left({1-\nu\over 2}  -\xi\right)
\log m
+\left({1-\nu\over 2} -\xi\right)\log 2
+\nu \log 2\pi\right.\\
&& \left.\hspace{20pt}\textstyle
-3{\mathbf g}\left({1+\nu\over 2}\right)
+{\mathbf g}\left( {1-\nu\over 2}\right)
+{\mathbf g}\left( \nu\right)
+{\mathbf g}\left( \xi\right)
+{\mathbf g}\left( 1-\xi\right)
\right.
\\
&& \left.\hspace{20pt}\textstyle
-{\mathbf g}\left({1+\nu\over 2}-\xi\right)
-{\mathbf g}\left({1+\nu\over 2}+\xi\right)
\right].
\end{eqnarray*}
Rescaling the integration variable $u \to y = b u$
and using (\ref{s_asym}) one has
\[
I(\alpha, \beta|s) \; \sim \; \frac{1}{b}\int\limits_{i{\mathbb R}}\!dy\;
\exp\left\{\frac{1}{b^2} {\mathbf f}(y,t,\xi,\nu)\right\},
\]
where
\begin{eqnarray}
\label{f}
{\mathbf f}(y,t,\xi,\nu)
& = & -
(1+\nu)\log 2- 2 t y \\
&+&
{\mathbf s}\left(\pi y + \frac{\pi}{4}(1-\nu-2\xi)\right)
+
{\mathbf s}\left(\pi y + \frac{\pi}{4}(1-\nu+2\xi)\right)\\
\nonumber
&-&
{\mathbf s}\left(\pi y + \frac{\pi}{4}(1+\nu-2\xi) + \frac{\pi}{2}\right)
-
{\mathbf s}\left(\pi y + \frac{\pi}{4}(1+\nu+2\xi) + \frac{\pi}{2}\right).
\nonumber
\end{eqnarray}
This integral can be approximated by its saddle point value.
The saddle point equation
\begin{eqnarray*}
\label{saddle}
\hspace*{-5mm}
{\rm e}^{2t}
=
\frac{
\sin \left(\pi y + \frac{\pi}{4}(1-\nu-2\xi)\right)
\sin \left(\pi y + \frac{\pi}{4}(1-\nu+2\xi)\right)}
{
\cos\left(\pi y + \frac{\pi}{4}(1+\nu-2\xi)\right)
\cos\left(\pi y + \frac{\pi}{4}(1+\nu+2\xi)\right)
}
& = &
\frac{
\cos\pi\xi + \sin\left(2\pi y - \frac{\pi\nu}{2}\right)
}
{
\cos\pi\xi - \sin\left(2\pi y + \frac{\pi\nu}{2}\right)
}
\end{eqnarray*}
yields two solutions
\begin{eqnarray}
\nonumber
\label{spoint:solution}
 \cos2\pi y_{s\pm}
 & = & \frac{1}{\cosh^2 t - \sin^2 {\pi\nu\over 2}}
 \left[
 \sinh^2t\,\cos\pi\xi\,\sin\frac{\pi\nu}{2} \right.
 \\
 &&\hspace{20pt}
 \left.
 \pm\  \cosh t\,\cos\frac{\pi\nu}{2}
  \sqrt{
\cos^2\frac{\pi\nu}{2}
+\left({\omega_{\rm \scriptscriptstyle B}^2\over 4}-1
\right)\sin^2\pi\xi}\,
 \right].
\end{eqnarray}
The appropriate solution could be in principle selected by a careful
analysis of the position of the contour with respect to
poles located on the real axis.
In the present calculations we have chosen $y_s=y_{s+}$ on
the basis of numerical checks of the final result instead.

The asymptotic takes the form
\[
\log I(\alpha, \beta|s) \sim {1\over b^2}{\mathbf f}\Big(y_s(t,\xi,\nu),t,\xi,\nu\Big),
\]
where $y_s(t,\xi,\nu)=y_{s\pm}(t,\xi,\nu)$.
The classical action calculated from the classical limit of
the quantum expression
$$
\langle V_\alpha(z_1) B^{ss}_\beta (\infty)\rangle\;
\stackrel{b\to 0}{\sim}\;
{\rm e}^{-{1\over b^2}S_{\rm cl}}
$$
reads
\begin{eqnarray*}
S_{\rm cl}&=&
\textstyle{1\over 2}\left({1-\nu\over 2}  -\xi\right)
\log m
+
\left({1\over 4}+{\nu^2\over 4}-{\xi^2\over 2}\right)\log|z_1-\bar z_1|
-\left({1-\nu\over 2} -\xi\right)\log 2
\\
&+&\log 2+
{\mathbf g}\left( \xi\right)
+{\mathbf g}\left( 1-\xi\right)
\\
&-&\nu \log \pi
+\textstyle
3{\mathbf g}\left({1+\nu\over 2}\right)
-{\mathbf g}\left( {1-\nu\over 2}\right)
-{\mathbf g}\left( \nu\right)
\\
&+&\textstyle
{\mathbf g}\left({1+\nu\over 2}-\xi\right)
+{\mathbf g}\left({1+\nu\over 2}+\xi\right)
+ 2 t y_s \\
&-&\textstyle
{\mathbf s}\left(\pi  y_s + \frac{\pi}{4}(1-\nu-2\xi)\right)
-
{\mathbf s}\left(\pi  y_s + \frac{\pi}{4}(1-\nu+2\xi)\right)\\
\nonumber
&+&\textstyle
{\mathbf s}\left(\pi  y_s + \frac{\pi}{4}(1+\nu-2\xi) + \frac{\pi}{2}\right)
+
{\mathbf s}\left(\pi  y_s + \frac{\pi}{4}(1+\nu+2\xi) + \frac{\pi}{2}\right),
\end{eqnarray*}
and therefore
\begin{eqnarray*}
\frac{\partial S_{\rm cl}}{\partial \xi}
& = &
-{1\over 2} \log m
+
\textstyle- \xi\log|z_1-\bar z_1|
 + \log 2
\\
\nonumber
& - &
\log\left[\frac{\Gamma(\xi)\Gamma\left(\frac{1+\nu}{2}-\xi\right)
}
{\Gamma(1-\xi)\Gamma\left(\frac{1+\nu}{2}+\xi\right)}\right]
+
\frac12\log\left[
\frac{
\cos 2\pi y_s - \sin\left(\pi\xi + \frac{\pi\nu}{2}\right)
}
{
\cos 2\pi y_s + \sin\left(\pi\xi - \frac{\pi\nu}{2}\right)
}
\right], \\
\frac{\partial S_{\rm cl}^{(1)}}{\partial \nu}
& = &\textstyle
-{1\over 4}\log m +{\nu\over 2}\log|z_1-\bar z_1|+
\log\left[
\frac{\Gamma\left(\frac{1+\nu}{2}\right)}{\Gamma(\nu)}
\right]
-{1\over 2}\log \pi\\
&+&\textstyle
{1\over 2}\log\left[\Gamma\left(\frac{1+\nu}{2}-\xi\right)
\Gamma\left(\frac{1+\nu}{2}+\xi\right)\right]
-{1\over 2}\log \cos\pi\left(\frac{1+\nu}{2}\right)
\\
&+&\textstyle
\frac14\log
\left[
\left(\cos 2\pi y_s - \sin\left(\pi\xi + \frac{\pi\nu}{2}\right)\right)
\left(\cos 2\pi y_s + \sin\left(\pi\xi - \frac{\pi\nu}{2}\right)\right)
\right],
\\
\frac{\partial S_{\rm cl}^{(1)}}{\partial t}
& = &
2y_s \; = \; \frac{1}{\pi}\,{\rm arcsin}\sqrt{1-\cos^2(2\pi y_s)}\ .
\end{eqnarray*}
These equations have to be compared with the equations (\ref{der:xi:bubo}),
(\ref{der:nu:bubo}) and (\ref{der:t:bubo}).
We have checked the corresponding equalities  by {\it
Mathematica 5.2} obtaining a perfect agreement in all three cases.

\section*{Acknowledgments}
This work was partially supported by the Polish State Research Committee (KBN) grant
no.\ 1 P03B 025 28.

\vskip 1mm
\noindent
The research of L.H. is supported by the Alexander von Humboldt Foundation scholarship.

\vskip 1mm
\noindent
Part of the ideas contained in this work emerged
during the 2005 Simons Workshop in Physics and Mathematics.
L.H.\  thanks the organizers of the Workshop for their warm hospitality,
which created at Stony Brook a highly stimulating environment
and for the financial support.

\section*{Appendix}
\renewcommand{\theequation}{{\rm A}.\arabic{equation}}
\setcounter{equation}{0}

We shall calculate
the length of the $j$-th boundary component
\begin{eqnarray*}
\ell_j=\int\limits_{L_j}\!dx \;{\rm e}^{\tilde\varphi(z,\bar z)/2}
& = &
\hspace*{-2mm}
\int\limits_{\rho(L_j)}\!\!\frac{|d\rho|}{\Im\,\rho}\,,
\end{eqnarray*}
in terms of the endpoints
$
\rho_j = \!\!\lim\limits_{x\to x_j +} \rho(x),\;\,
\rho_{j+1} = \!\!\!\!\lim\limits_{x\to x_{j+1} -} \rho(x).
$
For arbitrary two points $\rho_{j},\rho_{j+1}\in \mathbb{H}$ one can always
find
 an $SL(2,\mathbb{R})$
transformation $w(\rho)$ such that
$$
\Re w(\rho_{j+1})= -\Re w(\rho_{j})>0,
\hskip 5mm
\Im w(\rho_{j+1})= \Im w(\rho_{j}).
$$
It is then sufficient to calculate the hyperbolic length of the
curve $w\circ\rho(L_j)$ connecting
the points  $q_j=w(\rho_{j}),\; q_{j+1}=w(\rho_{j+1})$.
Let us note that the sign of the boundary geodesic curvature
depends on the location of $\rho(\cal M)$ with respect to $\rho(L_j)$,
so one can assume $\omega_j>0$ in the following.

For $0 < \omega_j\leqslant 2$ there are always two curves
with the geodesic curvature ${\omega_j\over 2}$
connecting points $q_j,q_{j+1}$ (see Fig.1).
These are arcs of two circles  with their centers on the imaginary axis and
their radii determined by the condition that the Euclidean distance
of the center from the real axis equals ${\omega_j\over 2}R$:
$$
R_\pm= {2\,\Im q_j\over 4-\omega_j^2 }\left(
\pm\omega_j
+\sqrt{(4-\omega^2_j)\beta^2 +4}
\right),
$$
where $\beta=\beta(q_j,q_{j+1})$
and
$$
\beta(z,w){=}
\frac{|z - w|}{2\sqrt{\Im\,z\Im\,w}}
$$
is an $SL(2,\mathbb{R})$-invariant.

\begin{figure}[h]
\centering

\includegraphics*[width=.65\textwidth]{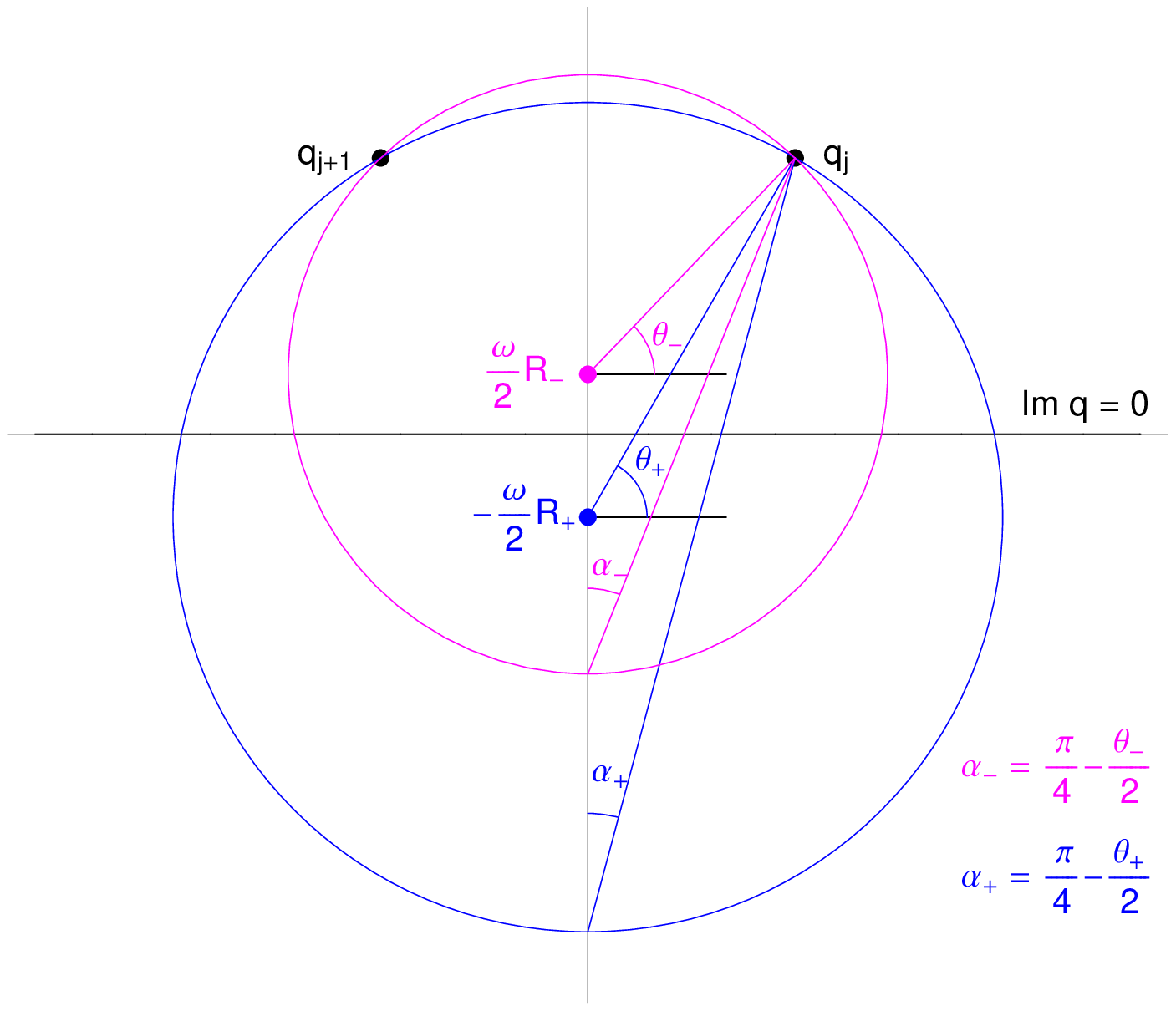}
\bigskip

\small {\bf Fig.~1}~ The geometry involved
in the determination of the boundary length, $\omega_j<2$.
\end{figure}

The hyperbolic lengths of the corresponding arcs
can be easily calculated
\begin{eqnarray*}
\ell_\pm
& = &
\int\limits_{\vartheta_\pm}^{\pi-\vartheta_\pm}
\!\!\frac{2 d\vartheta}{2\sin\vartheta \mp
 \omega_j}
 =
\frac{8}{\sqrt{4-\omega_j^2}}\, {\rm arctanh}
\left[\sqrt\frac{2\pm\omega_j}{2\mp\omega_j}
\tan\left(\frac{\pi}{4} - \frac{\theta_\pm}{2}\right)\right],
\end{eqnarray*}
where $\theta_\pm = {\rm Arg} (q_j\pm i {\omega_j\over 2}R_\pm)$.
It follows from Fig.1 that
\begin{eqnarray*}
\tan \left(\frac{\pi}{4} - \frac{\theta_\pm}{2}
\right)&=&
{{|q_{j+1}-q_j|\over 2} \over R_\pm +\sqrt{R^2_\pm -{|q_{j+1}-q_j|^2\over 4}}}
=
{(2\mp\omega_j) \beta  \over
2+\sqrt{(4-\omega^2_j)\beta^2 +4}}\ .
\end{eqnarray*}
Using the identity
$$
2 \,{\rm arctanh}\, {x\over 1 +\sqrt{1+x^2}}= {\rm arcsinh }\, x
$$
 one gets
\begin{equation}
\label{app:1}
\ell_\pm= \frac{4}{\sqrt{4-\omega_j^2}}\
{\rm arcsinh}\left[\sqrt{1-\left(\frac{\omega_j}{2}\right)^2}\
\beta(q_j,q_{j+1})\right].
\end{equation}
Since the lengths of both arcs are the same
one obtains $\ell_j=\ell_\pm$ which yields  formula (\ref{deromegafinal:1}).

In the case $\omega_j>2$ the points $q_j,q_{j+1}$
can be connected by
curves
with the geodesic curvature ${\omega_j\over 2}$
 if and only if
$
(\omega_j^2 -4)\beta^2\leqslant 4
$
(which is satisfied  for $\beta= \beta(q_j,q_{j+1})$ by construction).
One has two circles with the radii
$$
R_\pm= {2\Im q_j\over \omega^2-4 }\left(
\omega_j
\pm \sqrt{(4-\omega^2_j)\beta^2 +4}
\right),
$$
and four different arcs (see Fig.2).

\begin{figure}[h]
\centering

\includegraphics*[width=.65\textwidth]{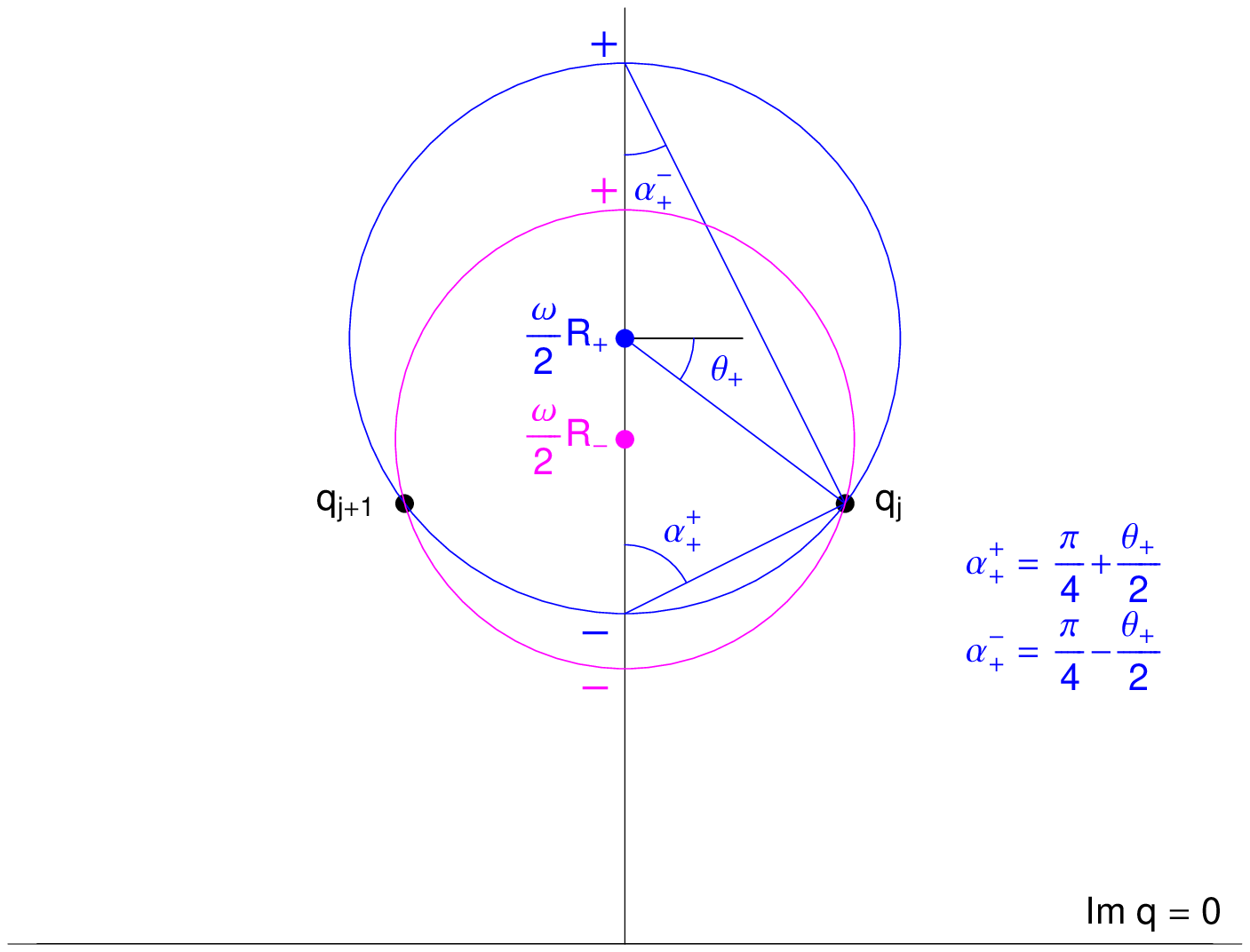}
\bigskip

\small {\bf Fig.~2}~ The geometry involved
in the determination of the boundary length, $\omega_j>2$.
\end{figure}

Let $\ell_\pm^+,\ell_\pm^-$ be the lengths of the upper,
and lower arcs of the circle with radius $R_\pm$, respectively.
Following essentially the same calculations as above
and using
\begin{eqnarray*}
2 \,{\rm arctan}\, {x\over 1 +\sqrt{1-x^2}}&=& {\rm arcsin }\, x,\\
2 \,{\rm arctan}\, {x\over 1 -\sqrt{1-x^2}}&=&\pi- {\rm arcsin }\, x,
\hskip 5mm
(0<x<1),
\end{eqnarray*}
 one gets
\begin{eqnarray}
\label{app:2}
\ell_-^+=\ell_+^-&=&
\frac{4}{\sqrt{\omega_j^2-4}}\
{\rm arcsin}\left[\sqrt{\left(\frac{\omega_j}{2}\right)^2-1}\
\beta(q_j,q_{j+1})\right],\\
\nonumber
\ell_-^-=\ell_+^+&=&
\frac{4}{\sqrt{\omega_j^2-4}}\
\left(\pi
-
\frac{4}{\sqrt{\omega_j^2-4}}\
{\rm arcsin}\left[\sqrt{\left(\frac{\omega_j}{2}\right)^2-1}\
\beta(q_j,q_{j+1})\right]
\right).
\end{eqnarray}

As the image $w\circ\rho (L_j)$ of the $j$-th boundary
component
is one of the four arcs, its hyperbolic
 length is  either $\ell_-^+$ or $\ell_-^-$.
On the other hand, the length of the $L_j$ depends
analytically on $\omega_j,$ so the formula for
$|\omega_j|>2$ should be an analytic continuation
of that   for $|\omega_j|<2$.
Taking this into account one finally gets
$\ell_j=\ell_-^+=\ell_+^-,$ what proves the formula (\ref{deromegafinal:2}).

\end{document}